\documentstyle[12pt]{article}

\newfont{\Bbb}{msbm10 scaled 1200}     %instead of eusb10
\newcommand{\mathbb}[1]{\mbox{\Bbb #1}}

\newcounter{apps}
\newcommand{\Ap}{\par\refstepcounter{apps}{\noindent\bf Appendix \arabic{apps}.}\bf}
\newcounter{prs}[section]
\newcommand{\Prop}{\par\refstepcounter{prs}%
{\noindent\bf Proposition \thesection.\arabic{prs}.}\it~}
\newcommand{\Ex}{\par{\bf Example}}
\newcounter{cors}
\newcommand{\Cr}{\par\refstepcounter{cors}{\noindent\bf Corollary \arabic{cors}.}~}
\newcounter{figs}
\newcommand{\fig}{\par\refstepcounter{figs}{\noindent\bf fig. \arabic{figs}}~}
\newcommand{\Rem}{\par{\noindent\bf Remark.}~}

\begin{document}

\begin{flushright}
ITEP-TH-70/01\\
\end{flushright}

\medskip

\begin{center}
\bigskip
{\large\bf Hopf algebra of ribbon graphs and renormalization}

\bigskip
\bigskip

Dmitry Malyshev

\bigskip

{\small\it Institute of Theoretical and Experimental Physics, Moscow\\
Moscow State University, Physics department\\
malyshev@gate.itep.ru}

\bigskip

{December 2001}

\bigskip
\bigskip
\bigskip

{\small{\bf Abstract}}
%\smallskip
\end{center}
{\small\it
Connes and Kreimer have discovered a Hopf algebra structure behind
renormalization of Feynman integrals.
We generalize the Hopf algebra to the case of ribbon graphs,
i.e. to the case of theories with matrix fields.
The Hopf algebra is naturally defined in terms of surfaces
corresponding to ribbon graphs.
As an example, we discuss renormalization of $\Phi^4$ theory and
the 1/N expansion.
}

\bigskip

\section{Introduction}
Connes and Kreimer have discovered a Hopf algebra of graphs \cite{CK1}.
This algebra underlies Bogolubov's recursive formula \cite{Bogolubov:1957}
for the renormalization of Feynman integrals.
The renormalization becomes more clear and natural from the
algebraic point of view.
The Hopf algebra of graphs is dual to the universal enveloping algebra of some Lie
algebra. This Lie algebra generates diffeomorphisms in the space of coupling
constants \cite{CK2,Kr,GMS}.

In this paper we construct the Hopf algebra of ribbon graphs.
Ribbon graphs appear in theories with matrix fields.
We cannot permute any two matrices in a product,
thus we are not allowed to permute any two edges of the vertex of
interaction.
If the Lagrangian has the form of a trace of matrices,
then there are only cyclic permutations of lines at any vertex
of the ribbon graph.
In the paper we consider a more general situation.
We assume that the lines attached to any vertex of the ribbon graph
can be divided into groups.
In any group cyclic permutations of the lines are allowed.
We also admit any permutations of the groups.
In this case the interaction has the form of a product of traces.

It is known that any ribbon graph corresponds to surfaces with
a boundary.
The ribbon graph can be drawn without self-intersections on any
corresponding surface.
There are infinitely many such surfaces but the surface with the
maximal Euler characteristic is unique.
We shall describe a one-to-one correspondence between ribbon graphs and
surfaces endowed with additional structures.
These structures have the form of cell decompositions.
In any cell decomposition the two dimensional cells are open
disks.
In our construction we use open spheres with holes instead of open disks.
This structure will be called a quasi-cell decomposition.
After that we construct a Hopf algebra of the surfaces with quasi-cell
decompositions.
This algebra immediately gives us
Bogolubov's recursive formula in theories with matrix
fields.
The Hopf algebra of surfaces and the corresponding Lie algebra may
show up in string theories.

The plan of the paper is as follows.
In Section 2 we define ribbon graphs and corresponding
surfaces with quasi-cell decompositions.
As an example, we consider the $\phi^4$ theory and
discuss the $\frac{1}{N}$ expansion.

In Section 3 we deal with the corresponding surfaces.
At first we introduce necessary definitions and essential
operations.
After that we define the Hopf algebra.

In Section 4 we consider dual objects to the Hopf
algebra.
The dual space is the space of linear functions on the Hopf
algebra.
This space is isomorphic to the universal enveloping algebra of a Lie
algebra.
The generators of the Lie algebra correspond to 1PI graphs \cite{CK1}.
After that we introduce the group of characters of the Hopf
algebra.
Any Quantum Field Theory can be viewed as a character
of the Hopf algebra of graphs.
Using the algebraic methods due to Connes and Kreimer \cite{CK1},
we obtain the renormalization of Feynman integrals.
The last section is devoted to the appendixes,
we give proofs and special constructions in this section.

\section{Ribbon graphs}

Any ribbon graph consists of vertices connected by lines.
Two vertices can be connected by several lines,
and a line can begin and finish at the same vertex.
In the latter case we say that the line makes a loop at the vertex.
Also a graph can be a disjoint union of connected graphs.
The vertices of any ribbon graph are called ribbon vertices.
In general, only cyclic permutations of lines are allowed in the ribbon vertices.

Let $\Gamma$ be a ribbon graph and let $C\in\Gamma$ be a ribbon vertex.
We denote by $\Lambda_C=\{l_1,l_2,\ldots l_n\}$ the set of all lines attached to the vertex.
If a line $l_i$ makes a loop at $C$, then the set $\Lambda_C$ contains this line twice:
$\Lambda_C=\{l_1,\ldots l'_i,\ldots l''_i,\ldots l_n\}$.
The order of elements of $\Lambda_C$ is the order in which the lines are attached to the
vertex.
The number of elements of $\Lambda_C$ is called the degree of $C\in\Gamma$ and is denoted
by ${\rm deg}_\Gamma(C)$.
If $p$ is the number of lines with one of the ends at $C$ and
$q$ is the number of lines with the both ends at $C$, then
$$
{\rm deg}_{\Gamma}(C)=p+2q.
$$

A ribbon vertex $C$ will be called simple
if the order of elements of $\Lambda_C$ is fixed
and only cyclic permutations are allowed.
For instance, $\{l_1,l_2,\ldots l_n\}\sim\{l_n,l_1,\ldots l_{n-1}\}$.
The corresponding interaction has the form of a trace.

A ribbon vertex $C$ will be called complex
if the set $\Lambda_C$ can be divided into subsets
$\Lambda_C^i,\;i=1,\ldots,m$ such that
$\Lambda_C=\{\Lambda_C^1,\Lambda_C^2,\ldots \Lambda_C^m\}$.
In each subset the order of elements is fixed and cyclic permutations are allowed.
In $\Lambda_C$ any permutations of the subsets are allowed:
$\{\ldots\Lambda_C^i,\ldots\Lambda_C^j,\ldots\}\sim
\{\ldots\Lambda_C^j,\ldots\Lambda_C^i,\ldots\}.$
%We shall
We denote by ${\rm deg}^i_{\Gamma}(C)$ the number of elements of the subset
$\Lambda_C^i$. We see that
$$
{\rm deg}_{\Gamma}(C)=\sum_{i=1}^m{\rm deg}^i_{\Gamma}(C).
$$
The corresponding interaction has the form of a product of traces.
\medskip
{\Ex. Consider the theory of scalar complex fields $\Phi(x)$
in 4-dimensional Euclidian space with the Lagrangian %(\ref{f4})
}

\begin{equation}\label{f4}%\ref{f4}
\begin{array}{lll}
L&=&L_0+L_{int},
\medskip\\
L_0&=&Tr(\partial\Phi^+\partial\Phi+m^2\Phi^+\Phi),
\medskip\\
L_{int}&=&g_1\:Tr[(\Phi^+\Phi)^2]
+g_2\:(Tr[\Phi^+\Phi])^2.
\end{array}
\end{equation}

Here we have denoted
$$
\partial\Phi^+\partial\Phi
=\sum^4_{\mu=1}\partial_\mu\Phi^+(x)\partial_\mu\Phi(x).
$$
The field $\Phi(x)$ is a complex $N\times N$ matrix.
We have
$$
\begin{array}{lll}
L_0&=&\partial{\Phi^+}^i_j\:\partial\Phi^j_i+m^2{\Phi^+}^i_j\:\Phi^j_i,
\medskip\\
L_{int}&=&g_1\:{\Phi^+}^i_j\:\Phi^j_k\:{\Phi^+}^k_l\:\Phi^l_i
+g_2\:({\Phi^+}^i_j\:\Phi^j_i)\:({\Phi^+}^k_l\:\Phi^l_k).
\end{array}
$$
There is a summation over all the repeating indices
$i,j,k,l=1,\ldots,N$.

The propagators in the theory $<{\Phi^+}^i_j\:\Phi^k_l>\sim\delta^i_l\:\delta^k_j$.
For brevity we do not discuss the momentum dependence here.
In Feynman diagrams the propagators have the form

\begin{picture}(400,90)
\put(120,40){\line(1,0){100}}
\put(92,35){\large ${\Phi^+}^{i}_{j}$}
\put(225,35){\large ${\Phi\,}^{j}_{i}$}
\put(40,5){\fig}
\end{picture}

It is convenient to represent the propagators as consisting of two threads with
arrows. We also put indices and signs at the ends.

\begin{picture}(400,90)
%first
\put(120,37){\vector(1,0){50}}
\put(170,37){\line(1,0){50}}
\put(220,43){\vector(-1,0){50}}
\put(170,43){\line(-1,0){50}}
\put(100,35){\Large $+$}
\put(122,45){\footnotesize $i$}
\put(122,27){\footnotesize $j$}
\put(225,35){\Large $-$}
\put(218,45){\footnotesize $i$}
\put(218,27){\footnotesize $j$}
\put(40,5){\fig}
\end{picture}

The sign $'+'$ corresponds to the field $\Phi^+$,
the sign $'-'$ corresponds to the field $\Phi$.
The incoming arrow corresponds to the upper index,
the outgoing arrow corresponds to the lower index.

The interaction vertices in the theory are

\begin{picture}(400,320)(25,0)
%frame
%left_up
\put(65,240){\line(1,0){90}}
\put(110,195){\line(0,1){90}}
\put(43,236){${\Phi^+}^{i}_{j}$}
\put(105,183){${\Phi\,}^{j}_{k}$}
\put(160,236){${\Phi^+}^{k}_{l}$}
\put(105,288){${\Phi\,}^{l}_{i}$}
\put(118,250){$g_1$}

%right_up
\put(245,240){\line(1,0){90}}
\put(290,195){\line(0,1){90}}
\put(223,236){${\Phi^+}^{i}_{j}$}
\put(285,183){${\Phi\,}^{l}_{k}$}
\put(340,236){${\Phi^+}^{k}_{l}$}
\put(285,288){${\Phi\,}^{j}_{i}$}
\put(298,250){$g_2$}

%left_down
\put(51,77){\large $+$}
\put(67,85){\footnotesize $i$}
\put(107,83){\vector(-1,0){22}}
\put(85,83){\line(-1,0){20}}
\put(67,68){\footnotesize $j$}
\put(65,77){\vector(1,0){22}}
\put(87,77){\line(1,0){20}}

\put(105,26){\large $-$}
\put(101,37){\footnotesize $j$}
\put(107,77){\vector(0,-1){22}}
\put(107,55){\line(0,-1){20}}
\put(116,36){\footnotesize $k$}
\put(113,35){\vector(0,1){22}}
\put(113,57){\line(0,1){20}}

\put(158,77){\large $+$}
\put(151,68){\footnotesize $k$}
\put(113,77){\vector(1,0){22}}
\put(135,77){\line(1,0){20}}
\put(151,85){\footnotesize $l$}
\put(155,83){\vector(-1,0){22}}
\put(133,83){\line(-1,0){20}}

\put(105,127){\large $-$}
\put(116,119){\footnotesize $l$}
\put(113,83){\vector(0,1){22}}
\put(113,105){\line(0,1){20}}
\put(101,119){\footnotesize $i$}
\put(107,125){\vector(0,-1){22}}
\put(107,103){\line(0,-1){20}}

\put(120,95){ $g_1$}

%right_down
\put(231,81){\large $+$}
\put(247,89){\footnotesize $i$}
\put(283,87){\vector(-1,0){18}}
\put(265,87){\line(-1,0){20}}
\put(247,72){\footnotesize $j$}
\put(245,81){\vector(1,0){22}}
\put(267,81){\line(1,0){22}}

\put(281,127){\large $-$}
\put(292,119){\footnotesize $j$}
\put(289,81){\vector(0,1){24}}
\put(289,105){\line(0,1){20}}
\put(277,119){\footnotesize $i$}
\put(283,125){\vector(0,-1){22}}
\put(283,103){\line(0,-1){16}}

\put(289,26){\large $-$}
\put(285,36){\footnotesize $l$}
\put(291,79){\vector(0,-1){24}}
\put(291,55){\line(0,-1){20}}
\put(300,36){\footnotesize $k$}
\put(297,35){\vector(0,1){22}}
\put(297,57){\line(0,1){16}}

\put(338,73){\large $+$}
\put(331,62){\footnotesize $k$}
\put(297,73){\vector(1,0){20}}
\put(317,73){\line(1,0){18}}
\put(331,81){\footnotesize $l$}
\put(335,79){\vector(-1,0){22}}
\put(313,79){\line(-1,0){22}}

\put(300,93){ $g_2$}

\put(40,5){\fig}
\end{picture}

The degree of the vertices is equal to four.
The left vertex is simple.
The right one is complex, indeed, the set of lines attached to the vertex consists of two
subsets with two lines in each subset.
\medskip

We see that the structure of vertices is rather
complicated in ribbon graphs.
We can consider two-dimensional objects instead of the vertices.
This enables us to describe
the different structures of vertices more efficiently.

Now we construct surfaces corresponding to ribbon graphs.
Let $\Gamma$ be a ribbon graph.
For simplicity, assume that all the vertices in $\Gamma$ are simple.
Let $C$ be a vertex of $\Gamma$, we associate to this vertex a polygon with
${\rm deg}_\Gamma(C)$ edges.
For example, any simple vertex with four attached lines corresponds to a square.

\begin{picture}(400,150)%vertex-square
%left_down
\put(51,77){\large $+$}
\put(67,85){\footnotesize $i$}
\put(107,83){\vector(-1,0){22}}
\put(85,83){\line(-1,0){20}}
\put(67,68){\footnotesize $j$}
\put(65,77){\vector(1,0){22}}
\put(87,77){\line(1,0){20}}

\put(105,26){\large $-$}
\put(101,37){\footnotesize $j$}
\put(107,77){\vector(0,-1){22}}
\put(107,55){\line(0,-1){20}}
\put(116,36){\footnotesize $k$}
\put(113,35){\vector(0,1){22}}
\put(113,57){\line(0,1){20}}

\put(158,77){\large $+$}
\put(151,68){\footnotesize $k$}
\put(113,77){\vector(1,0){22}}
\put(135,77){\line(1,0){20}}
\put(151,85){\footnotesize $l$}
\put(155,83){\vector(-1,0){22}}
\put(133,83){\line(-1,0){20}}

\put(105,127){\large $-$}
\put(116,119){\footnotesize $l$}
\put(113,83){\vector(0,1){22}}
\put(113,105){\line(0,1){20}}
\put(101,119){\footnotesize $i$}
\put(107,125){\vector(0,-1){22}}
\put(107,103){\line(0,-1){20}}

\put(120,95){ $g_1$}

\put(244,123){$i$}
\put(233,78){\large $+$}
\put(250,120){\vector(0,-1){42}}
\put(250,78){\line(0,-1){38}}
\put(238,30){ $j$}
\put(285,28){\large $-$}
\put(250,40){\vector(1,0){42}}
\put(292,40){\line(1,0){38}}
\put(333,30){$k$}
\put(335,78){\large $+$}
\put(330,40){\vector(0,1){42}}
\put(330,82){\line(0,1){38}}
\put(333,123){$l$}
\put(285,124){\large $-$}
\put(330,120){\vector(-1,0){42}}
\put(288,120){\line(-1,0){38}}
\put(194,76){\Large $\sim$}
\put(40,5){\fig}
\end{picture}

Any line at the vertex corresponds to an edge of the polygon.
We put indices, signs and arrows of the line on the
corresponding edge.
A line between two vertices is a common edge between the
corresponding polygons.

\begin{picture}(400,110)%connection
%left
%horizontal
\put(42,54){\footnotesize $i$}
\put(42,40){\footnotesize $n$}
\put(40,48){\line(1,0){28}}
\put(40,52){\line(1,0){28}}
\put(72,48){\line(1,0){56}}
\put(72,52){\line(1,0){56}}
\put(155,54){\footnotesize $k$}
\put(156,38){\footnotesize $l$}
\put(132,48){\line(1,0){28}}
\put(132,52){\line(1,0){28}}
%vertical
\put(60,23){\footnotesize $n$}
\put(75,23){\footnotesize $m$}
\put(68,25){\line(0,1){23}}
\put(72,25){\line(0,1){23}}
\put(61,71){\footnotesize $i$}
\put(75,71){\footnotesize $j$}
\put(68,52){\line(0,1){23}}
\put(72,52){\line(0,1){23}}
\put(117,23){\footnotesize $m$}
\put(135,23){\footnotesize $l$}
\put(128,25){\line(0,1){23}}
\put(132,25){\line(0,1){23}}
\put(120,71){\footnotesize $j$}
\put(134,71){\footnotesize $k$}
\put(128,52){\line(0,1){23}}
\put(132,52){\line(0,1){23}}

%right
\put(220,25){\line(1,0){100}}
\put(220,75){\line(1,0){100}}
\put(220,25){\line(0,1){50}}
\put(270,25){\line(0,1){50}}
\put(320,25){\line(0,1){50}}
\put(320,25){\line(0,1){50}}
\put(214,18){\footnotesize $n$}
\put(220,25){\circle*{3}}
\put(215,77){\footnotesize $i$}
\put(220,75){\circle*{3}}
\put(269,27){ $B$}
\put(266,18){\footnotesize $m$}
\put(270,25){\circle*{3}}
\put(269,64){ $A$}
\put(268,80){\footnotesize $j$}
\put(270,75){\circle*{3}}
\put(323,18){\footnotesize $l$}
\put(320,25){\circle*{3}}
\put(322,77){\footnotesize $k$}
\put(320,75){\circle*{3}}
{\fig \label{A}}
\end{picture}

We put indices near the ends of edges of the polygons.
We have only one index near any such point.
This property remains after the connection of the polygons.
For instance,
we have the index $j$ near the vertex $A$ and
the index $m$ near the vertex $B$.

In ribbon graphs the indices are associated with the threads.
Consequently the points at the ends of edges correspond to the threads in ribbon graphs.
We shall denote by indices the corresponding threads or points at the ends of edges.
For instance, we shall say 'the thread $k$' or 'the point $j$'.

If a line makes a loop at a vertex,
then we connect the edges of the corresponding polygon.

\begin{picture}(400,95)%index loop
%left
\put(62,35){\footnotesize $j$}
\put(62,20){\footnotesize $i$}
\put(60,28){\line(1,0){80}}
\put(60,32){\line(1,0){35}}
\put(135,35){\footnotesize $j$}
\put(135,20){\footnotesize $i$}
\put(105,32){\line(1,0){35}}
\put(98,36){\footnotesize $k$}
\put(95,46){\oval(16,28)[l]}
\put(105,46){\oval(16,28)[r]}
\put(95,60){\line(1,0){10}}
\put(100,45){\oval(15,21)}

\put(230,48){\circle{36}}
\put(230,48){\line(0,1){18}}
\put(228,20){\footnotesize $i$}
\put(230,30){\circle*{3}}
\put(224,40){\footnotesize $k$}
\put(230,48){\circle*{3}}
\put(228,71){\footnotesize $j$}
\put(230,66){\circle*{3}}

{\fig \label{thr}}
\end{picture}

Thus we can find the corresponding surface with a boundary for any ribbon graph with simple vertices.
The internal lines of the graph correspond to the internal edges on the surface,
the external lines correspond to the external edges
(the edges on the boundary of the surface).
%We leave edges of polygons on the surface.
Note that the surface is a CW-complex in this case.

Let us recall the definition of the CW-complex.
Any CW-complex is a union of cells.
We denote by $C^{(n)}$ an n-dimensional cell.
Zero-dimensional cells are points,
one-dimensional cells are open segments,
two-dimensional cells are diffeomorphic to open disks,
$n$-dimensional cells are diffeomorphic to open
$n$-dimensional balls for $n>0$.
Any $N$-dimensional CW-complex ${\mathcal{M}}_N$ is a union
\begin{equation}\label{CW}%\ref{CW}
{\mathcal{M}}_N=\bigcup_{n=0}^N\bigcup_{i_n}C^{(n)}_{i_n}
\end{equation}
such that the boundary of any $k$-dimensional cell
$C^{(k)}_{i_k}$ for all $k\in\{1,\:\ldots,\:N\}$
belongs to
$$
{\mathcal{M}}_{k-1}=\bigcup_{n=0}^{k-1}\bigcup_{i_n}C^{(n)}_{i_n}\subset{\mathcal{M}}_N,
$$
$$
fr\:C^{(k)}_{i_k}\subset{\mathcal{M}}_{k-1}.
$$
For simplicity, we assume that union (\ref{CW}) is finite.

\medskip
For example, suppose we have a square, the boundary of
any edge consists of the two points at the ends, the interior of the square has the
boundary consisting of the four edges.
Thus any square is a two-dimensional CW-complex.

We see that if we construct a surface by identifying, or gluing, edges of
polygons, then this surface is a CW-complex.
In this case we also say that the polygons give the cell decomposition of the
surface.
\medskip

{\Rem Let us consider fig. \ref{thr}. There is a closed thread inside the loop.
Hence we have a sum over index $k$ in the Feynman integral associated with this diagram.
The sum gives the factor of $N$.
Closed threads are also called index loops.
The fact that the index loops give the factors of $N$
is exploited in the $\frac{1}{N}$ expansion ('t Hooft \cite{'tHooft:1974jz}).
We see that the number of index loops
is an important characteristic of ribbon graphs.}
\medskip

We are going to find the objects on our surfaces which correspond to the index loops.
Let $S$ be a surface with a cell decomposition.
Let $j$ be a common end of the edges $l_1^j,\ldots,l_n^j\subset S$.
Since the surface is closed,
the point $j$ lies on the boundary if and only if at least one of the edges
$l_1^j,\ldots,l_n^j$ is external.
Indeed, if the vertex is external, then by
continuity there exists an external edge attached to this vertex.
On the other hand, if an edge is external, then the both ends of this edge
are external vertices.

Since the open threads end at external lines,
they correspond to points on the boundary of the corresponding surface.
The closed threads go along internal lines, hence
they correspond to internal points.
Thus index loops correspond to the internal points at the ends of edges.

Let $\Gamma$ be a ribbon graph without external lines
and let $S_\Gamma$ be the corresponding surface.
This surface has no boundary.
We can draw the graph on the surface without self intersections.
Indeed, we can put each vertex of the graph at the center of the corresponding polygon.
The lines of the graph intersect the corresponding edges on the surface.

We have noted that the polygons give a cell decomposition of the surface.
We see that the graph gives another cell decomposition of the surface.
By construction, the number of vertices of the graph equals the number of polygons,
i.e. two-dimensional cells.
The lines of the graph and the edges of the polygons are
one-dimensional cells, we have the same numbers of lines and edges.
One can view the threads of the ribbon graph as the 'boundaries' of faces, then
the number of faces is equal to the number of index loops.
The index loops correspond to the points at the ends of edges.
We see that the number of zero-dimensional cells of one decomposition is equal to
the number of two-dimensional cells of the other;
the numbers of one-dimensional cells are the same.
Such two cell decompositions are called dual.

We denote by $V$ the number of vertices of the graph $\Gamma$,
by $P$ the number of lines, and by $I$ the number of index loops.
Note that in our example the graph $\Gamma$ has no external lines
and the surface $S_\Gamma$ has no boundary.
We have $V$ polygons, $P$ edges and $I$ points at the ends of edges
on $S_\Gamma$.
Let $\chi$ be the Euler characteristic of $S_\Gamma$.
The Euler formula in this case is
$$
V-P+I=\chi.
$$
The last equation is very important in the $1/N$ expansion.
But if there are complex ribbon vertices, then this formula is not so simple.

Let $C$ be a complex vertex of a ribbon graph $\Gamma$.
Let $\Lambda_C$ be the set of lines attached to $C$.
Suppose that $\Lambda_C$ is divided  into $m$ subsets $\Lambda_C^i,\;i=1,\ldots,m$,
then we associate to the vertex a sphere with $m$ holes.
'A sphere with holes' means any manifold, topologically equivalent to a
sphere with holes.
Note that a sphere with one hole is equivalent to a polygon.
The subset $\Lambda_C^j$ corresponds to the $j$-th hole.
The boundary of this hole consists of
${\rm deg}^j_{\Gamma}(C)$ edges,
i.e. the number of edges in the $j$-th hole is equal to the number of lines of the
corresponding $j$-th subset.
A line between two vertices is a common edge between the two
corresponding spheres with holes.
We see that the surface corresponding to the graph $\Gamma$
has a decomposition into spheres with holes.
This decomposition resembles the cell decomposition
but open spheres with holes are not diffeomorphic to open disks.
Thus the decomposition on spheres with holes will be called a quasi-cell
decomposition.
The corresponding surface will be called a surface with
quasi-cell decomposition or simply a surface with decomposition.

Let $S_\Gamma$ be the surface with decomposition corresponding to a
ribbon graph $\Gamma$.
We denote by $V,\:P,\:I$ the numbers of vertices, lines, and index loops of $\Gamma$
respectively.
Then in $S_\Gamma$ we have $V$ spheres with holes,
$P$ internal edges and $I$ internal points at the ends of edges.
Let $m_k$ be the number of holes in the $k$-th sphere,
$k=1,\ldots,V$.
We assume that $m_k\geq 1$.
We can put $(m_k-1)$ edges on the $k$-th sphere
in order to obtain a cell decomposition of this sphere.
Such decomposition of all spheres yields a cell decomposition of $S_\Gamma$.
If $\chi$ is the Euler characteristic of $S_\Gamma$,
then the Euler formula is
$$
I-P-\sum_{k=1}^V(m_k-1)+V=\chi,
$$
or
$$
I-P+\sum_{k=1}^V(2-m_k)=\chi.
$$
The numbers of points and edges on the boundary cancel each other and we do not write
them in the formulas.

Denote by $\chi_{m_k}$ the Euler characteristic of the sphere with $m_k$
holes.
Since $\chi_{m_k}=2-m_k$, we have
\begin{equation}\label{Eulchar}
I-P+\sum_{k=1}^V\chi_{m_k}=\chi.
\end{equation}

The last formula is nontrivial from the point of view of graphs.
\medskip

{\Ex. Let us consider the $1/N$ expansion in the theory with Lagrangian (\ref{f4}).}

Let $\Gamma$ be a ribbon graph in this theory.
Suppose that we have $V_1$ vertices $g_1$ and $V_2$ vertices $g_2$.
The corresponding Euler characteristics are $\chi_1=1$ and $\chi_2=0$.
Let $P$ be the number of internal lines,
$R$ be the number of external lines,
$I$ be the number of index loops, and $\chi$ be the Euler characteristic of the corresponding
surface.
We associate to $\Gamma$ the number
$$X_\Gamma=N^I\cdot g_1^{V_1}\cdot g_2^{V_2}.$$

The numbers $V_1,\:V_2,\:P,\:R$ are not independent.
Indeed, we can put points near the internal ends of the lines.
Then there are four points near every vertex,
any internal line has two points,
any external line has one point.
The number of points is
\begin{equation}\label{points}%(ref{points})
4(V_1+V_2)=2P+R.
\end{equation}
We have
\begin{equation}\label{points2}%(ref{points2})
P=2(V_1+V_2)-\frac{R}{2}.
\end{equation}
Using (\ref{Eulchar})
$$
I-P+\sum_{k=1}^{V_1}\chi_1+\sum_{l=1}^{V_2}\chi_2=\chi,
$$
or
$$
I-P+V_1=\chi.
$$
We replace $P$ by $2(V_1+V_2)-\frac{R}{2}$ in the last formula and
obtain
$$
I-V_1-2V_2=\chi-\frac{R}{2}.
$$
If we assume that $g_1=\frac{g_1^0}{N}$ and $g_2=\frac{g_2^0}{N^2}$,
where $g_1^0$ and $g_2^0$ do not depend on $N$,
then
\begin{equation}\label{factor}
X_\Gamma=N^{I-V_1-2V_2}\cdot (g_1^0)^{V_1}\cdot (g_2^0)^{V_2}
\sim N^{\;\chi-R/2}.
\end{equation}
If the number of external lines is fixed, then $X_\Gamma$ depends only on topological
characteristics of the corresponding surface.
\medskip

We see that for a given ribbon graph we can construct the corresponding surface.
Now, assume that a surface consists of spheres with holes connected along the edges.
We can always find the ribbon graph corresponding to this surface.
Hence we have the one-to-one correspondence between the ribbon graphs and
the surfaces with decomposition into spheres with holes.

\section{Surfaces}

In this section we consider surfaces with a boundary.
The surfaces have decompositions into spheres with holes.
As above, 'sphere with holes' means any manifold topologically equivalent to
a sphere with holes.
We assume, that any sphere with holes has edges on its boundary.

Let $C$ be a sphere with holes.
The number of edges of $C$ is called the degree of $C$ and is denoted by
${\rm deg}(C)$.
Let $m$ be the number of holes of $C$.
We denote by
${\rm deg}^i(C)$ the number of edges on the boundary of the $i$-th hole,
$i=1,\ldots,m$.
We see that ${\rm deg}(C)=\sum_{i=1}^m{\rm deg}^i(C).$

Let $l_1$ and $l_2$ be different edges; they may belong either to one
sphere with holes or to different ones.
We can identify $l_1$ and $l_2$; the edge $l\sim l_1\sim l_2$ is internal.
We do not identify internal edges with any other edges,
i.e. one may identify only two (but not more) different external edges.
If we take several spheres with holes and identify some pairs of edges,
we obtain a surface.
We allow the surface to be a disjoint union of connected components.
We say that this surface has a decomposition into spheres with holes.
Such surfaces will be also called surfaces with decomposition.
In the previous section we have shown that such surfaces correspond to ribbon graphs.

Let $S$ be a surface with decomposition, and let $C\subset S$ be a sphere with holes.
We denote by ${\rm deg}_S(C)$ the degree of $C$ inside $S$ and define it by the formula
${\rm deg}_S(C)={\rm deg}(C)$.
Similarly, if $C$ has several holes, then ${\rm deg}_S^i(C)={\rm deg}^i(C)$.

Now we define the operation of extraction of a subsurface.
Let $S$ be a surface with decomposition,
and let $s$ be an open subsurface of $S$ (the points on the boundary of
$s$ do not belong to $s$).
Suppose that $s$ is a disjoint union of connected components $s^j,\;j=1,\ldots,J.$

There is a natural mapping of connected open surfaces to open spheres with holes.
We define the mapping by the following rule.
A surface with $h$ holes maps to a sphere with $h$ holes.
Now let $c^j$ be the open sphere with holes corresponding to $s^j$.
We can extract $s^j$ from $S$ and insert $c^j$ in place of $s^j$
in the way that the boundary of $s^j$ becomes the boundary of $c^j$.
We denote by $S/s$ the surface which we obtain after the extraction of all $s^j$ and
the insertion of the corresponding $c^j$.

Now we define the closure of $s$ with respect to $S$.
The closure of $s$ is denoted by $\bar{s}$.
We add edges on the boundary of $s$,
so that every sphere with holes inside $\bar{s}$ has the same degree as it has had in $S$.
In other words, let $C\subset S$ be a sphere with holes
and let $int(C)\subset s$.\footnote{We denote by $int(S)$
the interior of the surface $S$.}
We define $\bar{s}$ to be the closure of $s$
such that ${\rm deg}_{\bar{s}}(C)={\rm deg}(C)={\rm deg}_S(C)$.
If $C$ has $m$ holes, then
${\rm deg}^i_{\bar{s}}(C)={\rm deg}^i(C),\;i=1,\ldots,m.$

If $s$ is the empty surface $\emptyset$, then we define
$S/\emptyset=S$.
If $\bar{s}=S$, then we define $S/s=\emptyset$.

Also we use the following notation for the operation of extraction
$\varphi_s:\;S\rightarrow S/s$.
\medskip

{\Ex.}

\begin{picture}(400,110)%Extraction from nut.
%left
\put(60,43){\line(0,1){54}}
\put(60,70){\oval(50,54)}%[l]
\put(60,43){\circle*{3}}
\put(60,61){\circle*{3}}
\put(60,79){\circle*{3}}
\put(60,97){\circle*{3}}
\put(54,20){\large $S$}

%center
\put(188,46){\special{em: moveto}}
\put(196,50){\special{em: lineto}}
\put(199,53){\special{em: moveto}}
\put(203,61){\special{em: lineto}}
\put(204,65){\special{em: moveto}}
\put(204,74){\special{em: lineto}}
\put(203,78){\special{em: moveto}}
\put(199,86){\special{em: lineto}}
\put(196,89){\special{em: moveto}}
\put(188,93){\special{em: lineto}}
\put(184,94){\special{em: moveto}}
\put(175,94){\special{em: lineto}}
\put(171,93){\special{em: moveto}}
\put(163,89){\special{em: lineto}}
\put(160,86){\special{em: moveto}}
\put(156,78){\special{em: lineto}}
\put(155,74){\special{em: moveto}}
\put(155,65){\special{em: lineto}}
\put(156,61){\special{em: moveto}}
\put(160,53){\special{em: lineto}}
\put(163,50){\special{em: moveto}}
\put(171,46){\special{em: lineto}}
%angle
\put(172,48){\special{em: moveto}}
\put(175,54){\special{em: lineto}}
\put(176,56){\special{em: moveto}}
\put(179,62){\special{em: lineto}}
\put(181,62){\special{em: moveto}}
\put(184,56){\special{em: lineto}}
\put(185,54){\special{em: moveto}}
\put(188,48){\special{em: lineto}}
%vertical
\put(180,64){\special{em: moveto}}
\put(180,94){\special{em: lineto}}

\put(180,78){\circle*{3}}
\put(177,23){\large $s$}
\put(227,69){ $\rightarrow$}

%right
\put(298,46){\special{em: moveto}}
\put(306,50){\special{em: lineto}}
\put(309,53){\special{em: moveto}}
\put(313,61){\special{em: lineto}}
\put(314,65){\special{em: moveto}}
\put(314,74){\special{em: lineto}}
\put(313,78){\special{em: moveto}}
\put(309,86){\special{em: lineto}}
\put(306,89){\special{em: moveto}}
\put(298,93){\special{em: lineto}}
\put(294,94){\special{em: moveto}}
\put(285,94){\special{em: lineto}}
\put(281,93){\special{em: moveto}}
\put(273,89){\special{em: lineto}}
\put(270,86){\special{em: moveto}}
\put(266,78){\special{em: lineto}}
\put(265,74){\special{em: moveto}}
\put(265,65){\special{em: lineto}}
\put(266,61){\special{em: moveto}}
\put(270,53){\special{em: lineto}}
\put(273,50){\special{em: moveto}}
\put(281,46){\special{em: lineto}}
%angle
\put(282,48){\special{em: moveto}}
\put(285,54){\special{em: lineto}}
\put(286,56){\special{em: moveto}}
\put(289,62){\special{em: lineto}}
\put(291,62){\special{em: moveto}}
\put(294,56){\special{em: lineto}}
\put(295,54){\special{em: moveto}}
\put(298,48){\special{em: lineto}}

\put(284,20){ $C$}
{\fig \label{extr}}
\end{picture}

Here $s$ is an open subsurface of $S$ and
$C$ is the corresponding 'open sphere with holes'.
We see that $C$ is a sphere with one hole.

\begin{picture}(400,115)(0,-5)%Extracted.
%left
\put(70,43){\line(0,1){18}}
\put(70,70){\oval(50,54)}
\put(70,43){\circle*{3}}
\put(70,61){\circle*{3}}
\put(70,97){\circle*{3}}
\put(64,20){\large $S/s$}
%right
\put(200,70){\oval(50,54)[t]}
\put(190,70){\oval(30,54)[lb]}
\put(210,70){\oval(30,54)[rb]}
\put(190,43){\special{em: moveto}}
\put(200,61){\special{em: lineto}}
\put(210,43){\special{em: moveto}}
\put(200,61){\special{em: lineto}}
\put(200,61){\special{em: moveto}}
\put(200,97){\special{em: lineto}}
\put(190,43){\circle*{3}}
\put(210,43){\circle*{3}}
\put(200,61){\circle*{3}}
\put(200,79){\circle*{3}}
\put(200,97){\circle*{3}}
\put(197,23){\large $\bar{s}$}
{\fig \label{extred}}
\end{picture}

Here $S/s$ is the surface $S$ with extracted $s$,
the surface $\bar{s}$ is the closure of $s$ with respect to $S$.
Observe, that the number of internal edges on the surface $S$ is the same as
on $S/s$ and $\bar{s}$ together. Indeed, we have three internal edges on $S$,
one internal edge on $S/s$ and two internal edges on $\bar{s}$.
Also we have the conservation of the number of internal vertices,
there are two of them on $S$, one on $S/s$, and one on $\bar{s}$.
The third number, which is conserved, is the number of spheres with holes
minus the number of connected components of the surface $S$.

The corresponding ribbon graphs are planar
and we can draw them as the ordinary ones.

\begin{picture}(400,180)
 \special{em:linewidth 0.5pt}
%top
\put(40,150){\line(1,0){80}}
\put(80,150){\oval(40,40)}%[l]
\put(60,150){\circle*{3}}
\put(100,150){\circle*{3}}
\put(61,110){\large $\Gamma\sim S$}

%down-left
\put(40,80){\line(1,0){80}}
\put(80,60){\oval(30,40)}%[l]
\put(80,80){\circle*{3}}
\put(48,15){\large $\Gamma/\gamma\sim S/s$}

%down-right
\put(200,70){\line(1,0){80}}
\put(240,70){\oval(40,40)[t]}
\put(220,70){\circle*{3}}
\put(260,70){\circle*{3}}
\put(220,70){\line(0,-1){15}}
\put(260,70){\line(0,-1){15}}
\put(223,30){\large $\gamma\sim \bar{s}$}

{\fig \label{}}
\end{picture}

In graphs we have the conservation of the number of internal lines,
of the number of index loops, and of the number of vertices minus the number of connected
components of the graphs.
\medskip

We denote by ${\rm P}_S$ the number of internal edges on the surface $S$,
by ${\rm T}_S={\rm C}_S-con(S)$ the number of spheres with holes minus the number of connected
components of the surface $S$,
and by ${\rm I}_S$ the number of internal points at the ends of
internal edges on the surface $S$.

{\Rem\label{PTI} The operation of extraction conserves the
numbers ${\rm P}_S$, ${\rm T}_S$ and ${\rm I}_S$
$$
\begin{array}{ccccc}
{\rm P}_S&=&{\rm P}_{\bar{s}}&+&{\rm P}_{S/s}\,,
\medskip\\
{\rm T}_S&=&{\rm T}_{\bar{s}}&+&{\rm T}_{S/s}\,,
\medskip\\
{\rm I}_S&=&{\rm I}_{\bar{s}}&+&{\rm I}_{S/s}\;.
\end{array}
$$
We use the conservation of ${\rm P}_S$ in proofs by induction,
while the conservation of ${\rm I}_S$ is important for the $\frac{1}{N}$ expansion.}
\medskip

In the construction of the Hopf algebra Connes and Kreimer use 1PI (one particle
irreducible) graphs \cite{CK1}.
Similarly, we introduce 1PI surfaces.
A surface with decomposition is called 1PI if it is connected, contains at least 1 internal edge and
remains connected after the removal
of any internal edge.
Also we need the empty surface and the disjoint unions of 1PI surfaces.
We call these surfaces allowed.
Thus the set of allowed surfaces, denoted by $\{S_1,\:S_2\ldots\,\}$,
contains the empty surface, the 1PI surfaces and the disjoint unions of 1PI surfaces.
Note that this set is countable.

Let $S$ be an allowed surface.
We can restrict the operation of extraction to allowed open subsurfaces $s_0\subset S$.
An open surface $s$ is called allowed if the closure $\bar{s}$ is an allowed surface.
\medskip

The operation of extraction has the following properties
{\begin{enumerate}
\label{opex}
  \item Bijection.
Let $s_0\subset S$ be an open allowed subsurface.
Consider the set of closed allowed surfaces
$\{S_{1}:\;s_0\subset S_{1}\subset S\}$.
Since $s_0\subset S_{1}$, we can define the surface $S_{1}/s_0\subset S/s_0$.
Hence we have the mapping
$\varphi_{s_0}:\;\{S_{1}:\;s_0\subset S_{1}\subset S\}
\rightarrow\{S_1/s_0\}$.
Also it is not difficult to prove that the set of the surfaces $S_{1}/s_0$ is equal to the
set of allowed subsurfaces of $S/s_0$.
Thus we have the bijective mapping
$\varphi_{s_0}:\;\{S_{1}:\;s_0\subset S_{1}\subset S\}
\rightarrow\{S_{2}:\;S_{2}\subset S/s_0\}$.
  \item Composition law.
Let $s_1$ and $s_{12}$ be open subsurfaces of $S$ such that $s_1\subset s_{12}$.
We define $s_{12}/s_1$ as the interior of $\bar{s}_{12}/s_1$.
We denote by $s_2$ the surface ${s}_{12}/s_1$. Note that $s_2\subset S/s_1$.
The composition law is
\begin{equation}\label{comp}
(S/s_1)/s_2=S/s_{12}.
\end{equation}
In terms of mappings we have
$$
\varphi_{s_2}\circ\varphi_{s_1}=\varphi_{s_{12}}.
$$
The mappings are bijective,
thus for any given $s_1\subset S$ and $s_2\subset S/s_1$ there exists a unique
$s_{12}\subset S$ that satisfies (\ref{comp}).
\end{enumerate}}
These properties are necessary for the construction of a consistent Hopf algebra.
\medskip

Any Hopf algebra is a linear space endowed with the operations of a product, a unit,
a coproduct, a counit, and an antipode.
The Hopf algebra of surfaces is similar to the Hopf algebra of graphs
defined by Connes and Kreimer in the paper \cite{CK1}.

The basis in the linear space is labeled by all allowed surfaces.
It is convenient to designate the basis vectors by the same letters
$S_1,\:S_2\ldots$ as the corresponding surfaces.
We denote by ${\mathcal{H}}$ the linear space with the basis $\{S_1,\:S_2\ldots\,\}$
over the field ${\mathcal{F}}$ containing the field of rational numbers.
By definition, any vector $V\in{\mathcal{H}}$ is a linear combination
$$
V=\sum_{k=1}^{\infty}\lambda_k S_k,
$$
where $\lambda_k\in {\mathcal{F}}$ and $\lambda_k\neq 0$
for a finite number of indices $k\in \mathbb{N}$.
We use the notation $\mathbb{N}=\{1,2,\ldots\,\}$.

The product is a bilinear mapping
$m:\;{\mathcal{H}}\otimes{\mathcal{H}}\rightarrow{\mathcal{H}}$.
The basis in ${\mathcal{H}}\otimes{\mathcal{H}}$ is given by
$S_i\otimes S_j,\;i,j\in\mathbb{N}$.
We define the product on basis vectors as a disjoint union of the
corresponding surfaces:
\begin{equation}\label{m}
m(S_i\otimes S_j)=S_i\coprod S_j.
\end{equation}
We also use the notation
$m(S_i\otimes S_j)=S_i\cdot S_j$.
Obviously, the product $m$ is associative and commutative.

Further we denote by $\Gamma$ an arbitrary basis vector.
When we say that $\Gamma$ is a surface,
we mean the surface corresponding to the basis vector.
The letter $\Gamma$ is chosen to keep in mind
that the surfaces correspond to ribbon graphs.

Since
$$
\Gamma\cdot\emptyset=\emptyset\cdot\Gamma=\Gamma
$$
for any $\Gamma$,
the empty surface $\emptyset$ is the unit element with respect to the product $m$.

The unit is a linear mapping
$i:\;{\mathcal{F}}\rightarrow{\mathcal{H}}$.
We define it by the formula
\begin{equation}\label{i}
i(1)=\emptyset.
\end{equation}
Consequently $i(\lambda)=\lambda\emptyset$ for all $\lambda\in{\mathcal{F}}$.

The linear space ${\mathcal{H}}$ together with the product $m$
and the unit $i$ is an algebra.

\medskip
A linear space together with a coproduct and a counit is called a coalgebra.

The coproduct is a linear mapping
$\Delta:\;{\mathcal{H}}\rightarrow{\mathcal{H}}\otimes{\mathcal{H}}$.
We define the coproduct by the formula
\begin{equation}\label{coprod1}
\Delta\Gamma=\sum_{\gamma\subset\Gamma}\bar{\gamma}\otimes\Gamma/\gamma,
\end{equation}
the sum is over all allowed open subsurfaces $\gamma\subset\Gamma$.

\medskip

{\Ex. If $\Gamma=\emptyset$, then the only subset $\gamma\subset\Gamma$
is $\emptyset$ and the coproduct is $\Delta\emptyset=\emptyset\otimes\emptyset.$}
\medskip

For $\Gamma\neq\emptyset$ we shall also use two other formulas.
The first is
\begin{equation}\label{coprod2}
\Delta\Gamma=
\Gamma\otimes\emptyset+\emptyset\otimes\Gamma+
{\sum_{\gamma\subset\Gamma}}'\bar{\gamma}\otimes\Gamma/\gamma.
\end{equation}
The prime near the sum means that
the summation is over allowed open subsurfaces
$\gamma\subset\Gamma:\;\bar{\gamma}\neq\Gamma,\:\gamma\neq\emptyset$.
The second is
\begin{equation}\label{coprod3}
\Delta\Gamma=
\Gamma\otimes\emptyset+
{\sum^{\neq}_{\gamma\subset\Gamma}}\bar{\gamma}\otimes\Gamma/\gamma,
\end{equation}
where the sum is over all open $\gamma\subset\Gamma$ such that $\bar{\gamma}\neq\Gamma$,
note that $\gamma=\emptyset$ is allowed.

The coproduct is not cocommutative but coassociative
\begin{equation}\label{coas}
(\Delta\otimes id)\Delta\Gamma=(id\otimes\Delta)\Delta\Gamma,\;\; \forall\Gamma.
\end{equation}

The proof of the coassociativity is given in Appendix \ref{Ca}.

The counit is a linear mapping $\varepsilon:\;{\mathcal{H}}\rightarrow{\mathcal{F}}$.
We define the counit by the formula
\begin{equation}\label{counit}
\varepsilon(\Gamma)=\delta_{\Gamma,\emptyset}
=\left\{
  \begin{array}{lc}
    1\;\;\; if\;\; \Gamma=\emptyset; \\
    0\;\;\; if\;\; \Gamma\neq\emptyset.
  \end{array}
\right.
\end{equation}

Observe, that the coproduct and the counit map the spaces in the opposite directions
with respect to the product and the unit.

A linear space together with the structures of an algebra and of a coalgebra
is called a bialgebra if the following conditions are satisfied
$$
\begin{array}{l}
\Delta(\Gamma_1\cdot\Gamma_2)=\Delta\Gamma_1\cdot\Delta\Gamma_2,\\
\varepsilon(\Gamma_1\cdot\Gamma_2)=\varepsilon(\Gamma_1)\varepsilon(\Gamma_2).
\end{array}
$$
It is not difficult to prove using (\ref{coprod1}) and (\ref{counit})
that these conditions hold in the algebra of graphs.

The antipode is a homomorphism $S:\;{\mathcal{H}}\rightarrow{\mathcal{H}}$.
We define the antipode by the formula
\begin{equation}\label{anti1}
m(S\otimes id)\Delta(\Gamma)=i\varepsilon(\Gamma).
\end{equation}

We denote by $id$ the identity mapping.
The tensor product of mappings acts in the following way
$(S\otimes id)(\Gamma_1\otimes\Gamma_2)=S(\Gamma_1)\otimes\Gamma_2$.

In the case $\Gamma=\emptyset$ we have
$$
\begin{array}{l}
i\varepsilon(\emptyset)=i(1)=\emptyset;\\
m(S\otimes id)\Delta(\emptyset)=m((S\otimes id)(\emptyset\otimes\emptyset)
=S(\emptyset)\cdot\emptyset=S(\emptyset).
\end{array}
$$
Hence $S(\emptyset)=\emptyset$.

Let $\Gamma\neq\emptyset$. Then
$$
\begin{array}{lll}
i\varepsilon(\Gamma)=i(0)=0;
\medskip\\
m(S\otimes id)\Delta(\Gamma)
&=&m((S\otimes id)(\Gamma\otimes\emptyset+\emptyset\otimes\Gamma+
{\sum'_{\gamma\subset\Gamma}}\;\bar{\gamma}\otimes\Gamma/\gamma))
\medskip\\
&=&S(\Gamma)\cdot\emptyset+S(\emptyset)\cdot\Gamma+
{\sum'_{\gamma\subset\Gamma}}\;S(\bar{\gamma})\cdot\Gamma/\gamma
\medskip\\
&=&S(\Gamma)+\Gamma+{\sum'_{\gamma\subset\Gamma}}\;S(\bar{\gamma})\cdot\Gamma/\gamma\;.
\end{array}
$$
Consequently we have the recursive formula
\begin{equation}\label{anti2}
S(\Gamma)=-\Gamma-{\sum_{\gamma\subset\Gamma}}'\;S(\bar{\gamma})\cdot\Gamma/\gamma\;,
\end{equation}
or using (\ref{coprod3})
\begin{equation}\label{anti3}
S(\Gamma)=-{\sum^{\neq}_{\gamma\subset\Gamma}}\;S(\bar{\gamma})\cdot\Gamma/\gamma\;,
\end{equation}
The recursion is on the number of internal edges.
Indeed, since $\bar{\gamma}\neq\Gamma$ on the right hand side,
we have
$$P_{\bar{\gamma}}=P_\Gamma-P_{\Gamma/\gamma}<P_\Gamma.$$
The base of the recursion is $S(\emptyset)=\emptyset$,
because $P_\Gamma\geq 0$ and $P_\Gamma=0\Leftrightarrow\Gamma=\emptyset$ for
any allowed surface $\Gamma$.

{\Prop The antipode, defined recursively by (\ref{anti2}), is a homomorphism of ${\mathcal{H}}$.}

We have to prove that
$$S(\gamma_1\cdot\gamma_2)=S(\gamma_1)S(\gamma_2).$$
for any allowed $\gamma_1$ and $\gamma_2$.
The proof is given in Appendix \ref{Shom}.

Also the antipode $S$ satisfies
$$
m\circ(S\otimes {\rm id})\circ\Delta=i\circ\varepsilon=m\circ({\rm id}\otimes S)\circ\Delta
$$
or
$$
m(S\otimes {\rm id})(\Delta\Gamma)=i\varepsilon(\Gamma)=m({\rm id}\otimes
S)(\Delta\Gamma).
$$
This equation is called the axiom of the antipode.
The axiom is proved in Appendix \ref{Aa}.

There are two nontrivial facts in the construction of the Hopf algebra ${\mathcal{H}}$.
These facts are the coassociativity and the axiom of the antipode.
The proof of these axioms is based on the two properties of the operation of extraction.
These properties are the only required conditions.
Thus we can choose another set of allowed subsurfaces,
and if the properties are satisfied,
then we can construct another Hopf algebra.
%For example, let $\gamma$ be an open subsurface of $\Gamma$,
%the surface $\bar{\gamma}$ is the closure of $\gamma$ with respect to $\Gamma$.
%We say that $\gamma$ is an allowed subsurface of $\Gamma$
%if $\bar{\gamma}$ is an allowed surface and every connected component of
%$\bar{\gamma}$ has four external edges. Also the surface $\Gamma$ and the empty surface $\emptyset$ are allowed
%subsurfaces of $\Gamma$.
{\Ex.} Consider the theory with Lagrangian (\ref{f4}).
We construct the Hopf algebra of the surfaces corresponding
to the ribbon graphs in this theory.
We have two types of vertices in the theory.
The vertices $g_1$ correspond to squares,
the vertices $g_2$ correspond to cylinders with two edges on each boundary.
Also it is convenient to view the massive term as an interaction,
the vertices $m^2$ have two attached lines and
correspond to disks with two edges on the boundary.
Thus the corresponding surfaces consist of spheres with one or two holes
and two or four external edges.
Let $S$ be such surface.
Note that $S$ may have more than four external edges.
We can say that an open subsurface $s\subset S$ is allowed
if $\bar{s}$ is a disjoint union of 1PI surfaces
and, in addition, each 1PI surface has two or four external edges.
If we extract a 1PI surface with four (or two) external edges,
then we insert a sphere with four (or two) external edges,
this sphere has the same type as the other spheres with holes in the theory.
%Thus we say that a subsurface is allowed if it gives 1PI surfaces with two or four external edges after the extraction.
By definition, the surface itself and the empty surface are allowed subsurfaces.
We see that the set of surfaces
corresponding to ribbon graphs in the theory
is closed with respect to the operation of extraction of allowed subsurfaces.
It is not difficult to check that the properties of the operation of extraction
hold.
Thus we can construct the Hopf algebra of such surfaces.
This Hopf algebra will be denoted by ${\mathcal{H}}_4$.

The coproduct in ${\mathcal{H}}_4$ is
\begin{equation}\label{copr4}%\ref{copr4}
\Delta\Gamma=
\Gamma\otimes\emptyset+\emptyset\otimes\Gamma+
{\sum_{\gamma\subset\Gamma}}'\!{}_4\;\;\bar{\gamma}\otimes\Gamma/\gamma,
\end{equation}
the sum is over open $\gamma\subset\Gamma$ such that $\bar{\gamma}$ is a disjoint union of
1PI surfaces with two or four external edges.

\section{Functions on surfaces. Renormalization}

%At first we consider linear functions .
The linear functions on ${\mathcal{H}}$ form a linear space,
denoted by ${\mathcal{H}}^*$.
%this space is called dual to the space ${\mathcal{H}}$.

The set of allowed surfaces $\{S_1,S_2\ldots\}$ is formally a basis in
${\mathcal{H}}$.
The basis in the dual space is given by linear functions $S^i,\;i\in{\mathbb{N}}$
such that
$$
S^i(S_j)=\delta^i_j.
$$
It is convenient to denote
$S^i(S_j)=(S^i,S_j)$.
We can say that the function $S^i$ corresponds to the surface $S_i$,
and that the space ${\mathcal{H}}^{*}$ is also a linear space of surfaces.

We use the coproduct in ${\mathcal{H}}$ to define a product in ${\mathcal{H}}^*$.
The product is a mapping
$\mu:\;{\mathcal{H}}^*\otimes{\mathcal{H}}^*\rightarrow{\mathcal{H}}^*$.
The basis in ${\mathcal{H}}^*\otimes{\mathcal{H}}^*$
consists of the bilinear functions $S^i\otimes S^j$.
The pairing of $S^i\otimes S^j$ with the basis vectors of
${\mathcal{H}}\otimes{\mathcal{H}}$ is the following
$$
(S^i\otimes S^j,\: S_k\otimes S_l)=\delta^i_k\delta^j_l, \;\;\forall i,j,k,l\in{\mathbb{N}}.
$$
The product $\mu:\;S^i\otimes S^j\mapsto S^i* S^j$ is defined by the formula
$$
(S^i*S^j,\:S_k)=(S^i\otimes S^j,\:\Delta S_k), \;\;\forall i,j,k\in{\mathbb{N}},
$$
or
\begin{equation}\label{Pr}
S^i*S^j=\sum_k (S^i* S^j,\:S_k)S^k=\sum_k(S^i\otimes S^j,\:\Delta S_k)S^k.
\end{equation}

We have $(S^i\otimes S^j,\:\Delta S_k)\neq 0$ if there is an open subsurface
$s\subset S_k$ such that $\bar{s}=S_i$ and $S_k/s=S_j$.
Thus we get $S_j$ after the extraction of $S_i$ from $S_k$.

In order to find the product $S^i*S^j$ we define the operation of insertion
which is inverse to the operation of extraction.
Let $\gamma$ be a surface with $m$ holes and $R^i$ edges on the boundary of the $i$-th
hole, $i=1,\:\ldots,\:m$.
Also let $\Gamma$ be a surface and let $C$ be a sphere with holes inside $\Gamma$.
If $C$ has $m$ holes and ${\rm deg}^i_{\Gamma}(C)=R^i\;\;i=1,\:\ldots,\:m$,
then we can insert $int(\gamma)$ instead of $int(C)$.
We denote by $\gamma\circ\Gamma$ the new graph that contains $\gamma$ instead
of $C$.
It is probably better to denote the graph $\gamma\circ\Gamma$ by
$\gamma\circ^\alpha_c\Gamma$ in order to show that we insert $\gamma$ instead of $C$.
Sometimes, we can inset $\gamma$ in different inequivalent ways.
The index $'\alpha'$ denotes the way of insertion.
But for shortness we use the notation $\gamma\circ\Gamma$.

Let $\stackrel{\circ}{\gamma}=int(\gamma)$,
we see that $(\gamma\circ\Gamma)/\stackrel{\circ}{\gamma}=\Gamma$.

If $\gamma^*$ and $\Gamma^*$ are dual to $\gamma$ and $\Gamma$,
then we have
$$
(\gamma^** \Gamma^*,\:\Gamma')
=(\gamma^*\otimes \Gamma^*,\:\Delta \Gamma')
=(\gamma^*\otimes \Gamma^*,\:\gamma\otimes\Gamma'/\stackrel{\circ}{\gamma}+\cdots)
\neq 0.
$$

In order to define the product more precisely we introduce a new formula for the
coproduct.
Let $\Gamma$ be a surface.
Sometimes we can find different open subsurfaces $\gamma,\:\gamma'\subset\Gamma$
such that $\bar{\gamma}\sim\bar{\gamma'}$ and $\Gamma/\gamma\sim\Gamma/\gamma'$.
We denote by $N(\Gamma_1,\Gamma_2;\Gamma)$\label{symmet} the number of open subsurfaces
$\gamma\subset\Gamma$
such that $\bar{\gamma}\sim\Gamma_1$ and $\Gamma/\gamma\sim\Gamma_2.$
If there is no such subsurfaces, then $N(\Gamma_1,\Gamma_2;\Gamma)=0$.
{\Ex. Let $a$ be a surface and let $\Gamma=a^n,\;\Gamma_1=a^m,\;\Gamma_2=a^{n-m}$,
where $n\geq m$ and $n,m\in{\mathbb{N}}$.
Then $N(a^m,a^{n-m};a^n)=\left(_m^n\right)=\frac{n!}{m!(n-m)!}$.

The new formula for the coproduct is
$$
\Delta\Gamma=\sum_{\gamma\subset\Gamma}^{\not\sim}
N(\bar{\gamma},\Gamma/\gamma;\Gamma)\bar{\gamma}\otimes\Gamma/\gamma,
$$
the sign '$\not\sim$' means that
the sum is over open $\gamma\subset\Gamma$ that give different
(inequivalent)
$\bar{\gamma}$ and $\Gamma/\gamma$.

Now the product is
$$
S^i* S^j=\sum_k(S^i\otimes S^j,\:\Delta S_k)S^k
=\sum_k\sum_{\gamma\subset S^k}^{\not\sim}
N(\bar{\gamma},S_k/\gamma;S_k)(S^i\otimes S^j,\:\bar{\gamma}\otimes S_k/\gamma)S^k,
$$
and
$$
S^i* S^j=\sum_k N(S_i,\:S_j;\:S_k)S^k.
$$
We see that the sum is over all surfaces that appear after the insertion of
$S^i$ into $S^j$, multiplied by some symmetry factors.
We can introduce a new basis in order to get rid of the symmetry factors
$N(S_i,\:S_j;\:S_k)$.

The new basis is $\widetilde{S^i}=Z(S_i)S^i$,
where $Z(S_i)$ is the number if symmetries of the surface
$S_i$ (see Appendix \ref{sym}).
{\Ex.} Let $a$ be a surface without internal symmetries,
then $Z(a^n)=n!,\;n=1,2,\ldots$

Let $S^i$ and $S^j$ be 1PI surfaces.
In the new basis the product is
$$
\widetilde{S^i}*\widetilde{S^j}
=\sum^{\not\sim}_{s^i\circ s^j}\widetilde{(S^i\circ S^j\:)}+\widetilde{S^i}\cdot\widetilde{S^j},
$$
the sum is over all
inequivalent insertions of $S^i$ into $S^j$,
the surface $S^i\cdot S^j$ is a disjoint union of $S^i$ and $S^j$;
the notation $\widetilde{(S^i\circ S^j\:)}$ means
$Z(S_i\circ S_j)\;S^i\circ S^j$.

Since the coproduct '$\Delta$' is coassociative, the product '$*$'
is associative.
Thus we have the associative algebra ${\mathcal{H}}^*$
with respect to the product '$*$'.

Now we define the Lie algebra $\mathcal{L}$ with the basis labeled by the 1PI surfaces
$\widetilde{S^i}$.
The Lie bracket is
$$
\begin{array}{ccl}
[\widetilde{S^i},\widetilde{S^j}]_{*}
&=&\widetilde{S^i}*\widetilde{S^j}-\widetilde{S^j}*\widetilde{S^i}
\medskip\\
&=&\sum^{\not\sim}_{s^i\circ s^j}\widetilde{(S^i\circ S^j\:)}
+\widetilde{S^i}\cdot\widetilde{S^j}
-\sum^{\not\sim}_{s^j\circ s^i}\widetilde{(S^j\circ S^i\:)}
-\widetilde{S^j}\cdot\widetilde{S^i}
\medskip\\
&=&\sum^{\not\sim}_{s^i\circ s^j}\widetilde{(S^i\circ S^j\:)} -
\sum^{\not\sim}_{s^j\circ s^i}\widetilde{(S^j\circ S^i\:)}.
\end{array}
$$
We see that $[\widetilde{S^i},\widetilde{S^j}]_{*}\in{\mathcal{L}}$
since it is a sum of 1PI surfaces $\widetilde{(S^i\circ S^j\:)}$
and $\widetilde{(S^j\circ S^i\:)}$.
The Jacobi identity is satisfied, since the product '$*$' is associative.
The universal enveloping algebra $U({\mathcal{L}})\simeq{\mathcal{H}}^*$.
Thus the space of linear functions on ${\mathcal{H}}$
has the structure of the universal enveloping algebra of the Lie algebra
${\mathcal{L}}$.

Consider the Hopf algebra ${\mathcal{H}}_4$.
The dual Lie algebra ${\mathcal{L}}^{(4)}$ spans the vectors
corresponding to 1PI surfaces,
constructed of spheres with two or four external edges.
We represent ${\mathcal{L}}^{(4)}$ as a direct sum
\begin{equation}
\label{dec}
{\mathcal{L}}^{(4)}={\mathcal{L}}_1\oplus{\mathcal{L}}_2,
\end{equation}
where ${\mathcal{L}}_1$ spans the vectors
corresponding to 1PI surfaces with two or four external edges.

Let $\Gamma_1,\;\Gamma'_1\in{\mathcal{L}}_1$ and
$\Gamma_2,\;\Gamma'_2\in{\mathcal{L}}_2$.
Since we cannot insert the surfaces $\Gamma_2,\;\Gamma'_2$ into any other surfaces,
we have the following commutators
$$
\begin{array}{lll}
[\Gamma_2,\;\Gamma'_2]_*&=&0,
\medskip\\
{[}\Gamma_1,\;\Gamma_2{]}_*
&=&\sum^{\not\sim}_{s^i\circ s^j}\Gamma_1\circ\Gamma_2
\in{\mathcal{L}}_2,
\medskip\\
{[}\Gamma_1,\;\Gamma'_1{]}_*
&=&\sum^{\not\sim}_{s^i\circ s^j}\Gamma_1\circ\Gamma_1'
-\sum^{\not\sim}_{s^i\circ s^j}\Gamma_1'\circ\Gamma_1
\in{\mathcal{L}}_1,
\end{array}
$$
where we assume that all the surfaces are in the new basis.
We can write
$$
\begin{array}{lll}
[{\mathcal{L}}_2,\;{\mathcal{L}}_2]&=&0,
\medskip\\
{[}{\mathcal{L}}_1,\;{\mathcal{L}}_2{]}&\subset &{\mathcal{L}}_2,
\medskip\\
{[}{\mathcal{L}}_1,\;{\mathcal{L}}_1{]}&\subset &{\mathcal{L}}_1.
\end{array}
$$
We see that ${\mathcal{L}}_2$ resembles the Lie algebra of translations
and ${\mathcal{L}}_1$ resembles the Lie algebra of rotations.
The Poincar\'{e} group is a semi-direct product of translations by the action of
rotations.
If $G,\;G_1$ and $G_2$ are the Lie groups corresponding to the Lie algebras
${\mathcal{L}}^{(4)},{\mathcal{L}}_1,\;{\mathcal{L}}_2$,
then $G$ is a semi-direct product of $G_2$ by the action of $G_1$ \cite{CK1}.
Since $[{\mathcal{L}}_2,\;{\mathcal{L}}_2]=0$, the Lie group $G_2$ is abelian.

Now we proceed with the characters of ${\mathcal{H}}$.
A character is a linear function $\chi:\;{\mathcal{H}}\rightarrow{\mathcal{F}}$
such that
$$
\chi(\Gamma_1\cdot\Gamma_2)
=\chi(\Gamma_1)\chi(\Gamma_2),\;\;\;\;
\forall\;\Gamma_1,\:\Gamma_2\in{\mathcal{H}}.
$$

The set of characters has the structure of a group.
We define the product of characters $\chi_1$ and $\chi_2$
by the formula
\begin{equation}\label{prod1}%\ref{prod1}
\chi_1*\chi_2(\Gamma)=\chi_1\otimes\chi_2(\Delta\Gamma),\;\;\;
\forall\;\Gamma\in{\mathcal{H}},
\end{equation}
where $\chi_1\otimes\chi_2(\Gamma_1\otimes\Gamma_2)=\chi_1(\Gamma_1)\chi_2(\Gamma_2)$
is a character of ${\mathcal{H}}\otimes{\mathcal{H}}.$
In particular,
$$
\chi\otimes\chi(\Gamma_1\otimes\Gamma_2)
=\chi(\Gamma_1)\chi(\Gamma_2)=\chi(\Gamma_1\cdot\Gamma_2).
$$
The product $\chi_1*\chi_2$ is a character, since
$\Delta(\Gamma_1\cdot\Gamma_2)=\Delta\Gamma_1\cdot\Delta\Gamma_2$
and
$$
\begin{array}{lll}
\chi_1*\chi_2(\Gamma_1\cdot\Gamma_2)
&=&\chi_1\otimes\chi_2(\Delta\Gamma_1\cdot\Delta\Gamma_2)\\
&=&\chi_1\otimes\chi_2(\Delta\Gamma_1)
\;\chi_1\otimes\chi_2(\Delta\Gamma_2)\\
&=&\chi_1*\chi_2(\Gamma_1)\;\chi_1*\chi_2(\Gamma_2).
\end{array}
$$
The unit of the group is the counit of ${\mathcal{H}}$.
We have
$$
\begin{array}{lll}
\varepsilon*\chi(\Gamma)
&=&\varepsilon\otimes\chi(\Delta\Gamma)
=\varepsilon\otimes\chi(\emptyset\otimes\Gamma+\cdots)\\
&=&\varepsilon(\emptyset)\;\chi(\Gamma)=\chi(\Gamma)\:,
\end{array}
$$
and similarly
$$
\chi*\varepsilon(\Gamma)=\chi(\Gamma)\:.
$$
The inverse character $\chi^{-1}=\chi\circ S$
or $\chi^{-1}(\Gamma)=\chi(S(\Gamma))$.
We have
$$
\begin{array}{lll}
\chi^{-1}*\chi(\Gamma)
&=&\chi^{-1}*\chi(\Delta\Gamma)
=\chi\otimes\chi((S\otimes id)\Delta\Gamma)\\
&=&\chi(m(S\otimes id)\Delta\Gamma)
=\delta_{\emptyset,\Gamma}\chi(\emptyset)
=\delta_{\emptyset,\Gamma}=\varepsilon(\Gamma)\:.
\end{array}
$$
The axiom of the antipode gives
$$
\chi*\chi^{-1}(\Gamma)=\varepsilon(\Gamma)\:.
$$

{\Ex. Exponentiation.}
Consider the Lie algebra ${\mathcal{L}}^{(4)}$ dual to
the Hopf algebra ${\mathcal{H}}_{4}$.
We have the decomposition
${\mathcal{L}}^{(4)}={\mathcal{L}}_1\oplus{\mathcal{L}}_2$
defined by (\ref{dec}).
Let $a\in{\mathcal{L}}_2$ be a surface such that the symmetry factor
$Z(a)=1$.
We define $a^k$ to be a disjoint union of $k$ surfaces $a$,
by definition $a^0=\emptyset.$
Assume that $a$ has no allowed subsurfaces except for the empty one
and the surface itself.
Then the coproduct of $a^n$ is
$$
\Delta a^n=\sum_{j=0}^n\left(_j^n\right)a^j\otimes a^{n-j}.
$$
The product $a^k*a^l$ in the dual algebra is
$$
a^k*a^l=(a^k\otimes a^l,\Delta a^{k+l})a^{k+l}=\left(^{k+l}_l\right)a^{k+l},
$$
or
$$
a^k*a^l=\frac{(k+l)!}{k!\:l!}a^{k+l}.
$$
In the new basis $\widetilde{a^n}=n!a^n$. We have
$$
\widetilde{a^k}*\widetilde{a^l}=\widetilde{a^{k+l}}.
$$
We may complete the Hopf algebra by the formal series and define
$$
\widetilde{e^{a}}=\sum_{n=0}^{\infty}\frac{1}{n!}\widetilde{a^n}.
$$
Note that
$$
\widetilde{e^{a}}=\sum_{n=0}^{\infty}\frac{1}{n!}\widetilde{a^n}
=\sum_{n=0}^{\infty}a^n.
$$
Now we prove that $\widetilde{e^{a}}$ is a character.
Let $\Gamma_1,\;\Gamma_2\in{\mathcal{H}}$, then
$$
(\widetilde{e^{a}},\Gamma_1\cdot\Gamma_2)
=\sum_{n=0}^{\infty}\frac{1}{n!}(\widetilde{a^n},\Gamma_1\cdot\Gamma_2)
=\sum_{n=0}^{\infty}\delta_{a^n,\Gamma_1\cdot\Gamma_2}\;.
$$
If $\Gamma_1=a^k$ and $\Gamma_2=a^l$ for some $k$ and $l$, then
$(\widetilde{e^{a}},\Gamma_1\cdot\Gamma_2)=1$, else
$(\widetilde{e^{a}},\Gamma_1\cdot\Gamma_2)=0$.
Similarly, we have
$$
(\widetilde{e^{a}},\Gamma_1)(\widetilde{e^{a}},\Gamma_2)
=\sum_{i=0}^{\infty}\delta_{a^i,\Gamma_1}
\sum_{j=0}^{\infty}\delta_{a^j,\Gamma_2}\;.
$$
As before, we see that
if $\Gamma_1=a^k$ and $\Gamma_2=a^l$ for some $k$ and $l$, then
$(\widetilde{e^{a}},\Gamma_1)(\widetilde{e^{a}},\Gamma_2)=1$, else
$(\widetilde{e^{a}},\Gamma_1)(\widetilde{e^{a}},\Gamma_2)=0$.
Consequently
$$
(\widetilde{e^{a}},\Gamma_1\cdot\Gamma_2)
=(\widetilde{e^{a}},\Gamma_1)(\widetilde{e^{a}},\Gamma_2),
$$
end we see that $\widetilde{e^{a}}$ is a character.

Let ${\mathcal{R}}$ be a ring containing the field ${\mathcal{F}}$.
Suppose that ${\mathcal{R}}={\mathcal{R}}_-\oplus{\mathcal{R}}_+$ and
$$
\begin{array}{cc}
{\mathcal{R}}_- + {\mathcal{R}}_-\subset{\mathcal{R}}_-&\;\;\;
{\mathcal{R}}_+ + {\mathcal{R}}_+\subset{\mathcal{R}}_+\\
{\mathcal{R}}_-\cdot{\mathcal{R}}_-\;\subset{\mathcal{R}}_-&\;\;\;
{\mathcal{R}}_+\cdot{\mathcal{R}}_+\;\subset{\mathcal{R}}_+
\end{array}
$$
{\Ex. The ring of Laurent series.
The space ${\mathcal{R}}_-$ consists of polynomials in $1/z$ without constant
terms
$$f_-=\sum_{i=1}^n a_i\left(\frac{1}{z}\right)^i,\;f_-\in{\mathcal{R}}_-.$$
The space ${\mathcal{R}}_+$ consists of series
$$f_+=\sum_{i=0}^{\infty} a_iz^i,\;f_+\in{\mathcal{R}}_+.$$}

We denote by
$P_-:\;{\mathcal{R}}\rightarrow{\mathcal{R}}_-$
the projector on ${\mathcal{R}}_-$.

Let $F$ be a character, $F:\;{\mathcal{H}}\rightarrow{\mathcal{R}}$.
Note, that Feynman integrals in dimensional regularization are such characters.
Indeed, if $\Gamma_1$ and $\Gamma_2$ are graphs-surfaces and $\Gamma_1\cdot\Gamma_2$ is
their disjoint union, then the Feynman integral
$F(\Gamma_1\cdot\Gamma_2)=F(\Gamma_1)F(\Gamma_2)$.
Also in dimensional regularization Feynman integrals are Laurent series
in the parameter of regularization.

We say that $F_+$ is a positive character if
$F_+:\;{\mathcal{H}}\rightarrow{\mathcal{R}}_+$.
We say that $F_-$ is a negative character if
$(1-P_-)F_-(\Gamma)=\delta_{\emptyset,\Gamma}$,
i.e. $F_-(\emptyset)=1$ and $P_-F_-(\Gamma)=F_-(\Gamma),\;\forall\Gamma\neq\emptyset$.
\medskip
{\Prop For any character $F$ there exists a positive character $R$ such that
$R=C* F$, where $C$ is a negative character.
Characters $C$ and $R$ depend on $F$.
The product '$*$' is defined by (\ref{prod1}).}

In terms of Feynman integrals, $C$ is the counterterm,
$R$ is the renormalized value of the Feynman integral $F$.

We define $C$ recursively.
$$
C(\emptyset)=1.
$$
Assume that we have found $C(\gamma)$ for all surfaces $\gamma$ with the number of
internal lines $P_\gamma\leq N$.
Let $\Gamma$ be a surface with $P_\Gamma=N+1$.
We have two conditions on $C$
\begin{equation}\label{C}
\left\{
  \begin{array}{lc}
    P_-(C* F)(\Gamma)=0; \\
    P_-C(\Gamma)=C(\Gamma).
  \end{array}
\right.
\end{equation}
By the definition of the product
$$
\begin{array}{lcl}
(C* F)(\Gamma)&=&(C\otimes F)(\Delta\Gamma)
\medskip\\
&=&(C\otimes F)(\Gamma\otimes\emptyset+\emptyset\otimes\Gamma+
{\sum'_{\gamma\subset\Gamma}}\;\bar{\gamma}\otimes\Gamma/\gamma)
\medskip\\
&=&C(\Gamma)\cdot F(\emptyset)+C(\emptyset)\cdot F(\Gamma)
+{\sum'_{\gamma\subset\Gamma}}\;C(\bar{\gamma})\cdot F(\Gamma/\gamma).
\end{array}
$$
Applying conditions (\ref{C}) we have
$$
P_{-}(C* F)(\Gamma)
=C(\Gamma)+P_{-}
\left(F(\Gamma)
+{\sum_{\gamma\subset\Gamma}}'\;C(\bar{\gamma})\cdot F(\Gamma/\gamma)
\right)
=0
$$
and
\begin{equation}\label{BR}
C(\Gamma)=-P_{-}
\left(F(\Gamma)
+{\sum_{\gamma\subset\Gamma}}'\;C(\bar{\gamma})\cdot F(\Gamma/\gamma)
\right).
\end{equation}
Since in the sum on the right hand side
$\gamma\subset int(\Gamma)$ and $\gamma\neq int(\Gamma)$,
the number of internal edges
$P_\gamma<P_\Gamma=N+1$, i.e. $P_\gamma\leq N$.
Thus we have defined $C(\Gamma)$.
Formula (\ref{BR}) is also called Bogolubov's recursion.

{\Prop Let $F$ be a character, then the function $C$ defined by (\ref{BR})
is a character,
i.e. $C(\Gamma_1\cdot\Gamma_2)=C(\Gamma_1)C(\Gamma_2),\;\forall\Gamma_1,\:\Gamma_2$.}

The proof is given in Appendix \ref{Nc}.

{\Cr The function $R$ as a product of characters $C$ and $F$ is a character.
Also we see that $C$ is a negative character and $R$ is a positive one.}

{\Ex.} Renormalizations of the theory $\Phi^4$ with Largangian (\ref{f4}).
\noindent Let $\Gamma$ be a ribbon graph.
The Feynman integral corresponding to $\Gamma$ is divergent if
$$
\omega_\Gamma=dL_\Gamma-2P_\Gamma\geq 0,
$$
where $d=4$ is the dimension,
$L_\Gamma=P_\Gamma-V_\Gamma+1$ is the number of independent integrations,
$P_\Gamma$ is the numbers of internal lines, and $V_\Gamma$ is the number of vertices.
Formula (\ref{points}) yields
$$
4V_\Gamma=2P_\Gamma+R_\Gamma,
$$
where $R_\Gamma$ is the number of external lines.
Hence
$$
\omega_\Gamma=4(P_\Gamma+1)-4V_\Gamma-2P_\Gamma
=4-R_\Gamma.
$$
Thus the divergent integrals correspond to the graphs with $R_\Gamma\leq 4$.
If $\Gamma\neq\emptyset$ and $R_\Gamma>4$,
then $C(\Gamma)=0$.
The renormalized Feynman integral
$$
R(\Gamma)=(C\otimes F)(\Delta\Gamma)
=\sum_{\gamma\subset\Gamma}C(\bar{\gamma})F(\Gamma/\gamma)
=\sum_{\gamma\subset\Gamma}\!{}_4\;C(\bar{\gamma})F(\Gamma/\gamma)
$$
the last sum is over open $\gamma\subset\Gamma$ such that $\bar{\gamma}$
is a disjoint union of 1PI surfaces with two or four external edges,
also we have $\bar{\gamma}=\emptyset$ and $\bar{\gamma}=\Gamma$ in the sum.
We see that for the renormalization it suffices to construct the Hopf algebra
${\mathcal{H}}_4$.

Now we find the structure of the renormalization in the $\frac{1}{N}$ expansion.
Graphs-surfaces which give the renormalization of the vertices $g_1$ contain four external edges
and have the topology of disks with handles,
hence by (\ref{factor}) we have
$$
\Delta g_1^{(ren)}\sim N^{-\chi-2}=N^{-1-2g},
$$
where $g$ is the number of handles.
The surfaces corresponding to the vertices $g_2$ have the topology of
cylinders with handles,
consequently the renormalization
$$
\Delta g_2^{(ren)}\sim N^{-\chi-2}=N^{-2-2g}.
$$
Finally, we have
$$
\begin{array}{lll}
g_1^{(ren)}&=&\frac{a_1}{N}+\frac{a_3}{N^3}+\frac{a_5}{N^5}+\cdots\;,
\medskip\\
g_2^{(ren)}&=&\frac{a_2}{N^2}+\frac{a_4}{N^4}+\frac{a_6}{N^6}+\cdots\;.
\end{array}
$$

\section{Appendices}

{\Ap \label{Ca} Coassociativity.}

We have to prove
\begin{equation}\label{coass2}
(id\otimes\Delta)\Delta\Gamma=(\Delta\otimes id)\Delta\Gamma,\;\;
\forall\Gamma.
\end{equation}
Using the left hand side:
$$
(id\otimes\Delta)\Delta\Gamma
=(id\otimes\Delta)
{\sum_{\gamma\subset\Gamma}}\bar{\gamma}\otimes\Gamma/\gamma
$$
$$
={\sum_{\gamma_1\subset\Gamma}}\;\;{\sum_{\gamma_2\subset\Gamma/\gamma_1}}
\bar{\gamma}_1\otimes\bar{\gamma}_2\otimes(\Gamma/\gamma_1)/\gamma_2.
$$
We use the second property of the operation of extraction
in order to find $\gamma_{12}\subset\Gamma$ such that
$(\Gamma/\gamma_1)/\gamma_2=\Gamma/\gamma_{12}$ and
$\bar{\gamma}_{12}/\gamma_1=\gamma_2$.
Since the operation of extraction is bijective we can change the sum over
$\gamma_2\subset\Gamma/\gamma_1$ to the sum over
$\gamma_{12}:\;\gamma_1\subset\gamma_{12}\subset\Gamma$.
After that we change the order of summation.
$$
(id\otimes\Delta)\Delta\Gamma
={\sum_{\gamma_1\subset\Gamma}}\;\;{\sum_{\gamma_{12}:\;\gamma_1\subset\gamma_{12}\subset\Gamma}}
\bar{\gamma}_1\otimes\bar{\gamma}_{12}/\gamma_1\otimes\Gamma/\gamma_{12}
$$
$$
={\sum_{\gamma_{12}\subset\Gamma}}\;\;{\sum_{\gamma_1\subset\bar{\gamma}_{12}}}
\bar{\gamma}_1\otimes\bar{\gamma}_{12}/\gamma_1\otimes\Gamma/\gamma_{12}
=(\Delta\otimes id){\sum_{\gamma_{12}\subset\Gamma}}
\bar{\gamma}_{12}\otimes\Gamma/\gamma_{12}
=(\Delta\otimes id)\Delta\Gamma\;.
$$
Thus the axiom is proved.
\medskip

{\Ap \label{Shom} The antipode as a homomorphism of the Hopf algebra.}

The antipode is defined by
\begin{equation}\label{anti1ap}
m(S\otimes id)\Delta(\Gamma)=i\varepsilon(\Gamma).
\end{equation}
The coproduct of $\Gamma$ is
$$
\Delta\Gamma=
\Gamma\otimes\emptyset+
{\sum^{\neq}_{\gamma\subset\Gamma}}\bar{\gamma}\otimes\Gamma/\gamma,
$$
where the sum is over all open $\gamma\subset\Gamma$ such that
$\bar{\gamma}\neq\Gamma$;
note that the number of internal edges
$P_{\bar{\gamma}}<P_\Gamma$ in this case.
Formula (\ref{anti1ap}) gives the recursion
\begin{eqnarray}
\label{anti2ap}
S(\emptyset)&=&\emptyset\nonumber,\\
S(\Gamma)&=&-{\sum^{\neq}_{\gamma\subset\Gamma}}S(\bar{\gamma})\cdot\Gamma/\gamma
\Gamma\neq\emptyset.
\end{eqnarray}
We must prove that
$S(\Gamma_1\cdot\Gamma_2)=S(\Gamma_1)\cdot S(\Gamma_2),\;\forall\Gamma_1,\;\Gamma_2$.

We prove it by induction on the number of internal edges.
If $P_{\Gamma_1\cdot\Gamma_2}=0$, then $\Gamma_1=\Gamma_2=\emptyset$
and $S(\Gamma_1\cdot\Gamma_2)=S(\Gamma_1)S(\Gamma_2)=\emptyset$.

Assume that we have proved the assertion for all
$\gamma_1$ and $\gamma_2$ such that $P_{\gamma_1\cdot\gamma_2}\leq N$.
Now we prove this for $\Gamma_1$ and $\Gamma_2$ such that
$P_{\Gamma_1\cdot\Gamma_2}=N+1$.
If $\Gamma_1=\emptyset$ or $\Gamma_2=\emptyset$
then the proof is trivial.
Thus we assume that $\Gamma_1\neq\emptyset$ and
$\Gamma_2\neq\emptyset$.
Also since ${\mathcal{H}}$ is a bialgebra,
$\Delta(\Gamma_1\cdot\Gamma_2)=\Delta\Gamma_1\cdot\Delta\Gamma_2$.
By definition (\ref{anti1ap}) we have
$$
m(S\otimes id)(\Delta\Gamma_1\cdot\Delta\Gamma_2)=0,
$$
thus
\begin{equation}\label{}
\begin{array}{lll}
0&=&m(S\otimes id)
(\Gamma_1\otimes\emptyset+{\sum^{\neq}_{\gamma_1\subset\Gamma_1}}
\bar{\gamma}_1\otimes\Gamma_1/\gamma_1)
(\Gamma_2\otimes\emptyset+{\sum^{\neq}_{\gamma_2\subset\Gamma_2}}
\bar{\gamma}_2\otimes\Gamma_2/\gamma_2)
\medskip\\
&=&S(\Gamma_1\cdot\Gamma_2)
+\sum^{\neq}_{\gamma_1\subset\Gamma_1}S(\Gamma_2\cdot\bar{\gamma}_1)\cdot\Gamma_1/\gamma_1
+\sum^{\neq}_{\gamma_2\subset\Gamma_2}S(\Gamma_1\cdot\bar{\gamma}_2)\cdot\Gamma_2/\gamma_2
\medskip\\
&&+\sum^{\neq}_{\gamma_1\subset\Gamma_1}\sum^{\neq}_{\gamma_2\subset\Gamma_2}
S(\bar{\gamma}_1\cdot\bar{\gamma}_2)\cdot\Gamma_1/\gamma_1\cdot\Gamma_2/\gamma_2
\medskip\\
&=&S(\Gamma_1\cdot\Gamma_2)+S(\Gamma_2)\cdot(-S(\Gamma_1))
+S(\Gamma_1)\cdot(-S(\Gamma_2))+(-S(\Gamma_1))\cdot(-S(\Gamma_2))
\medskip\\
&=&S(\Gamma_1\cdot\Gamma_2)-S(\Gamma_1)\cdot S(\Gamma_2),
\end{array}
\end{equation}
here we have used equations (\ref{anti2ap}) and the inequalities
$P_{\Gamma_1\cdot\bar{\gamma}_2},\:P_{\Gamma_2\cdot\bar{\gamma}_1},\:
P_{\bar{\gamma_1}\cdot\bar{\gamma}_2}<P_{\Gamma_1\cdot\Gamma_2}=N+1$.
We see that
$S(\Gamma_1\cdot\Gamma_2)=S(\Gamma_1)\cdot S(\Gamma_2)$.

\medskip
{\Ap \label{Aa} Axiom of the antipode.}

At first we discuss some general facts concerning surfaces
and after that we prove the axiom.

Let $\Gamma$ be a surface with decomposition.
As always we denote by $P_\Gamma$ the number of internal edges on $\Gamma$.
Let $\{\gamma_0,\:\gamma_1,\ldots,\:\gamma_{n+1}\}$ be a set of open surfaces
such that $\gamma_i\neq\gamma_j\;\forall\:i\neq j$
and
$\emptyset=\gamma_{n+1}\subset\gamma_n\subset\ldots\subset\gamma_0=int(\Gamma)$.
Assume that for all $i=0,\:\ldots,\:n$ there is no $\gamma\subset\Gamma$ such that
$\gamma\neq\gamma_i,\,\gamma_{i+1}$ and $\gamma_i\subset\gamma\subset\gamma_{i+1}$,
in this case the set $\{\gamma_0,\:\gamma_1,\ldots,\:\gamma_{n+1}\}$
will be called a maximal decreasing sequence of subsurfaces of $\Gamma$.
Note that $n\leq P_\Gamma$, thus the sequence is finite.

Consider the mapping
$\varphi_{\gamma_{i+1}}:\;\bar{\gamma}_i\rightarrow\bar{\gamma}_i/\gamma_{i+1}$.
Since the mapping is bijective and the set
$\{\gamma:\;\gamma_i\subset\gamma\subset\bar{\gamma}_{i+1}\}$ consists of the two elements
$\{\bar{\gamma}_i,\;\bar{\gamma}_{i+1}\}$,
we have
$\{\widetilde{\gamma}:\;\widetilde{\gamma}\subset\bar{\gamma}_i/\gamma_{i+1}\}
=\{\emptyset,\;\bar{\gamma}_i/\gamma_{i+1}\}$.
Consequently
$$
\Delta(\bar{\gamma}_i/\gamma_{i+1})
=\bar{\gamma}_i/\gamma_{i+1}\otimes\emptyset
+\emptyset\otimes\bar{\gamma}_i/\gamma_{i+1}
$$
and $S(\bar{\gamma}_i/\gamma_{i+1})=-\bar{\gamma}_i/\gamma_{i+1}$,
in particular $S(\bar{\gamma}_n)=-\bar{\gamma}_n$.

Also we shall use the following fact.
If $\{\gamma_0,\:\gamma_1,\ldots,\:\gamma_{n+1}\}$
is a maximal decreasing sequence of subsurfaces of $\Gamma$,
then
\begin{equation}\label{seq}
S(\bar{\gamma}_n)\cdot\bar{\gamma}_{n-1}/\gamma_n
-\bar{\gamma}_n\cdot S(\bar{\gamma}_{n-1}/\gamma_n)
=-\bar{\gamma}_n\cdot\bar{\gamma}_{n-1}/\gamma_n
+\bar{\gamma}_n\cdot\bar{\gamma}_{n-1}/\gamma_n
=0.
\end{equation}
We denote by $M_\Gamma$ the number of elements in the longest
decreasing sequence.
Note that $M_\Gamma\leq P_\Gamma+1$.
\medskip

The axiom of the antipode is
$$
m(S\otimes id)\Delta\Gamma
=i\varepsilon(\Gamma)
=m(id\otimes S)\Delta\Gamma.
$$
The left equation is used for the definition of the antipode
\begin{equation}\label{antdef}
\begin{array}{lll}
S(\emptyset)&=&\emptyset,\\
S(\Gamma)&=&-\Gamma-{\sum}'_{\gamma\subset\Gamma}S(\bar{\gamma})\cdot\Gamma/\gamma.
\end{array}
\end{equation}
We prove the right equation
$$
m(id\otimes S)\Delta\Gamma=i\varepsilon(\Gamma).
$$
If $\Gamma=\emptyset$, then
$$
m(id\otimes S)(\emptyset\otimes\emptyset)
=\emptyset\cdot S(\emptyset)
=\emptyset=i\varepsilon(\emptyset).
$$
Assume that $\Gamma\neq\emptyset$.
We prove the assertion by induction.
It is convenient to denote $\bar{\gamma}_0=\Gamma$
and $\Sigma=m(id\otimes S)\Delta\bar{\gamma}_0$.

The first step of the induction is
$$
\begin{array}{lll}
\Sigma&=&m(id\otimes S)\Delta\bar{\gamma}_0
=\bar{\gamma}_0+S(\bar{\gamma}_0)
+{\sum}'_{\gamma_1\subset\bar{\gamma}_0}S(\bar{\gamma}_1)\cdot\bar{\gamma}_0/\gamma_1
\medskip\\
&=&(-1){\sum}'_{\gamma_1\subset\bar{\gamma}_0}
\left(S(\bar{\gamma}_1)\cdot\bar{\gamma}_0/\gamma_1
-\bar{\gamma}_1\cdot S(\bar{\gamma}_0/\gamma_1)\right),
\end{array}
$$
we have used recursive formula (\ref{antdef}) for $S(\bar{\gamma}_0)$.

The second step is
$$
\begin{array}{llll}
\Sigma&=&
(-1)\;\;{\sum}'_{\gamma_1\subset\bar{\gamma}_0}
\!\!\left(\{
         -\bar{\gamma}_1
         -{\sum}'_{\gamma_2\subset\bar{\gamma}_1}S(\bar{\gamma}_2)\cdot\bar{\gamma}_1/\gamma_2\}
\cdot\bar{\gamma}_0/\gamma_1\right.\medskip\\
&&\left.+\bar{\gamma}_1\cdot\{
         \bar{\gamma}_0/\gamma_1
         +{\sum}'_{\gamma'\subset\bar{\gamma}_0/\gamma_1}
              S(\bar{\gamma}')\cdot(\bar{\gamma}_0/\gamma_1)/\gamma'
         \}\right)
\medskip\\
&=&(-1)^2\left({\sum}'_{\gamma_1\subset\bar{\gamma}_0}{\sum}'_{\gamma_2\subset\bar{\gamma}_1}
S(\bar{\gamma}_2)\cdot\bar{\gamma}_1/\gamma_2\cdot\bar{\gamma}_0/\gamma_1\right.\medskip\\
&&-\left.
{\sum}'_{\gamma_1\subset\bar{\gamma}_0}{\sum}'_{\gamma'\subset\bar{\gamma}_0/\gamma_1}
\bar{\gamma}_1\cdot S(\bar{\gamma}')\cdot(\bar{\gamma}_0/\gamma_1)/\gamma'
\right).
\end{array}
$$
Similarly, as in the proof of the coassociativity, we find $\gamma'_1$ such that
$(\bar{\gamma}_0/\gamma_1)/\gamma'=\bar{\gamma}_0/\gamma'_1$ and
$\bar{\gamma}'=\bar{\gamma}'_1/\gamma_1$.
After that we change the order of the summation in the second term
$$
{\sum}'_{\gamma_1\subset\bar{\gamma}_0}{\sum}'_{\gamma'\subset\bar{\gamma}_0/\gamma_1}
\bar{\gamma}_1\cdot S(\bar{\gamma}')\cdot(\bar{\gamma}_0/\gamma_1)/\gamma'
=
{\sum}'_{\gamma'_1\subset\bar{\gamma}_0}{\sum}'_{\gamma_1\subset\bar{\gamma}'_1}
\bar{\gamma}_1\cdot S(\bar{\gamma}'_1/\gamma_1)\cdot\bar{\gamma}_0/\gamma'_1.
$$
If we change the notations $\gamma_1\rightarrow\gamma_2$
and $\gamma'_1\rightarrow\gamma_1$, then
$$
\begin{array}{llll}
\Sigma&=&(-1)^2
{\sum}'_{\gamma_1\subset\bar{\gamma}_0}{\sum}'_{\gamma_2\subset\bar{\gamma}_1}
\left(S(\bar{\gamma}_2)\cdot\bar{\gamma}_1/\gamma_2\cdot\bar{\gamma}_0/\gamma_1
-\bar{\gamma}_2\cdot S(\bar{\gamma}_1/\gamma_2)\cdot\bar{\gamma}_0/\gamma_1
\right)\medskip\\
&=&(-1)^2
{\sum}'_{\gamma_1\subset\bar{\gamma}_0}{\sum}'_{\gamma_2\subset\bar{\gamma}_1}
\left(S(\bar{\gamma}_2)\cdot\bar{\gamma}_1/\gamma_2
-\bar{\gamma}_2\cdot S(\bar{\gamma}_1/\gamma_2)
\right)\cdot\bar{\gamma}_0/\gamma_1.
\end{array}
$$
Assume that after the $n$-th step we have
$$
\begin{array}{llll}
\Sigma=(-1)^n\;
{\sum}'_{\gamma_1\subset\bar{\gamma}_0}\cdots{\sum}'_{\gamma_n\subset\bar{\gamma}_{n-1}}
&\left(S(\bar{\gamma}_n)\cdot\bar{\gamma}_{n-1}/\gamma_n
-\bar{\gamma}_n\cdot S(\bar{\gamma}_{n-1}/\gamma_n)\right)\:\cdot
\medskip\\
&\cdot\;\bar{\gamma}_{n-2}/\gamma_{n-1}\cdot\ldots\cdot\bar{\gamma}_0/\gamma_1.
\end{array}
$$
We repeat the reasonings of the second step and obtain
$$
\begin{array}{llll}
{\sum}'_{\gamma_n\subset\bar{\gamma}_{n-1}}
\left(S(\bar{\gamma}_n)\cdot\bar{\gamma}_{n-1}/\gamma_n
-\bar{\gamma}_n\cdot S(\bar{\gamma}_{n-1}/\gamma_n)\right)
\medskip\\
={\sum}'_{\gamma_n\subset\bar{\gamma}_{n-1}}{\sum}'_{\gamma_{n+1}\subset\bar{\gamma}_{n}}
\left(S(\bar{\gamma}_{n+1})\cdot\bar{\gamma}_{n}/\gamma_{n+1}
-\bar{\gamma}_{n+1}\cdot S(\bar{\gamma}_{n}/\gamma_{n+1})\right)
\cdot\bar{\gamma}_{n-1}/\gamma_n.
\end{array}
$$
Thus after the $(n+1)$-st step we have
$$
\begin{array}{llll}
\Sigma=(-1)^{n+1}\;
{\sum}'_{\gamma_1\subset\bar{\gamma}_0}\cdots{\sum}'_{\gamma_{n+1}\subset\bar{\gamma}_{n}}
&\left(S(\bar{\gamma}_{n+1})\cdot\bar{\gamma}_{n}/\gamma_{n+1}
-\bar{\gamma}_{n+1}\cdot S(\bar{\gamma}_{n}/\gamma_{n+1})\right)\:\cdot
\medskip\\
&\cdot\;\bar{\gamma}_{n-1}/\gamma_{n}\cdot\ldots\cdot\bar{\gamma}_0/\gamma_1.
\end{array}
$$
Observe that the sums are over decreasing sequences of subsurfaces of $\Gamma$.
If $n+2=M_\Gamma$, then after the $n$-th step we have the sum only over maximal
decreasing sequences
$\{\gamma_0,\:\gamma_1,\ldots,\:\gamma_{n+1}\}$,
where $\gamma_{n+1}=\emptyset$ and $\gamma_{0}=int(\Gamma)$.
\newline
Since we have
$S(\bar{\gamma}_n)\cdot\bar{\gamma}_{n-1}/\gamma_n
-\bar{\gamma}_n\cdot S(\bar{\gamma}_{n-1}/\gamma_n)=0$
in this case, we conclude that
$$
\begin{array}{lll}
\Sigma&=(-1)^n\;
\sum'_{\{\gamma_1,\ldots,\:\gamma_{n}\}}\!\!\!\!
&\left(S(\bar{\gamma}_n)\cdot\!\bar{\gamma}_{n-1}/\gamma_n
-\bar{\gamma}_n\!\cdot\! S(\bar{\gamma}_{n-1}/\gamma_n)\right)\\
&&\;\;\;\cdot\,\bar{\gamma}_{n-2}/\gamma_{n-1}\cdot\ldots\cdot\bar{\gamma}_0/\gamma_1
\medskip\\
&=0,
\end{array}
$$
thus the axiom is proved.

\medskip
{\Ap \label{sym} Symmetries of surfaces.}

Let $\Gamma$ be a surface with decomposition,
and let $P_\Gamma$ be the number of internal edges.
Let us enumerate the internal edges by the indices $i=1,\:\ldots,\:P_\Gamma$.
We denote by
$\Gamma_I$ the surface $\Gamma$ with the enumerated internal edges.
The order of elements of the set $I$ is important.
Let $j:\;\Gamma_I\rightarrow\Gamma_{\sigma(I)}$ be a continuous automorphism of
$\Gamma$.
This automorphism does not move the boundary of
$\Gamma$ and maps internal edges $I$ to $\sigma(I)$,
where $\sigma(I)$ is a permutation of elements of $I$.
We say that two automorphisms
$j_1:\;\Gamma_I\rightarrow\Gamma_{\sigma_1(I)}$ and
$j_2:\;\Gamma_I\rightarrow\Gamma_{\sigma_2(I)}$ are equivalent if
$\sigma_1(I)=\sigma_2(I)$.
We say that surfaces are equivalent $\Gamma'_I\sim\Gamma''_J$ if there is an
automorphism $j:\;\Gamma'_I\rightarrow\Gamma_J''$, i.e. $\Gamma'=\Gamma''$ and
$J={\sigma(I)}$.
For brevity, sometimes we shall omit the subscript $\sigma(I)$.
We denote by ${\rm Z}(\Gamma)$ the number of continuous automorphisms
$j:\;\Gamma_I\rightarrow\Gamma_{\sigma(I)}$ up to the equivalence.
We also say that $j$ is a symmetry of $\Gamma$ and that ${\rm Z}(\Gamma)$ is the number of
the symmetries of $\Gamma$.

\medskip
Consider the following examples of surfaces without boundary.
{\Ex {\bf ~1.} Let $\Gamma$ be a sphere with two internal edges.
We can permute these edges, thus there are two automorphisms
$j_1:\;\Gamma_{\{1,2\}}\rightarrow\Gamma_{\{1,2\}}$ and
$j_2:\;\Gamma_{\{1,2\}}\rightarrow\Gamma_{\{2,1\}}$.
In this case ${\rm Z}(\Gamma)=2$.}
{\Ex {\bf ~2.} Let $\Gamma$ be a sphere with a 'square'.
Then we can rotate this 'square', thus there are four
automorphisms

\noindent
$j_1:\;\Gamma_{\{1,2,3,4\}}\rightarrow\Gamma_{\{1,2,3,4\}}$,
$j_2:\;\Gamma_{\{1,2,3,4\}}\rightarrow\Gamma_{\{2,3,4,1\}}$,\\
$j_3:\;\Gamma_{\{1,2,3,4\}}\rightarrow\Gamma_{\{3,4,1,2\}}$ and
$j_4:\;\Gamma_{\{1,2,3,4\}}\rightarrow\Gamma_{\{4,1,2,3\}}$.}
\par\noindent
The number of symmetries ${\rm Z}(\Gamma)=4$.
\medskip

Let $\Gamma_1$ and $\Gamma_2$ be surfaces such that $N(\Gamma_1,\Gamma_2;\Gamma)\neq
0$.
As above $N(\Gamma_1,\Gamma_2;\Gamma)$ is
the number of open subsurfaces
$\gamma\subset\Gamma$
such that $\bar{\gamma}\sim\Gamma_1$ and $\Gamma/\gamma\sim\Gamma_2.$
By definition, we have $n=N(\Gamma_1,\Gamma_2;\Gamma)$ open subsurfaces
$\gamma_i\subset\Gamma$ such that $\bar{\gamma_i}\sim\Gamma_1$ and
$\Gamma/\gamma_i\sim\Gamma_2$, $i=1,\:\ldots,\:n$.
Hence we have automorphisms
$j_i:\;\Gamma/\gamma_1\rightarrow\Gamma/\gamma_i$.
One can find the corresponding automorphisms
$j_i':\;\Gamma\rightarrow\Gamma$.
The automorphism $j_i'$ takes the subsurface $\gamma_1$ to the subsurface $\gamma_i$.
All other automorphisms
$j:\;\Gamma\rightarrow\Gamma$ do not change the boundary of $\gamma_1$.
They are either the internal automorphisms of $\gamma_1$ or the automorphisms of $\Gamma$ that do not
change $\gamma_1$.
The last ones do not depend on $\gamma_1$ and remain in $\Gamma/\gamma_1$.
Sometimes new symmetries appear after the extraction of a subsurface.
Let $\gamma$, $\Gamma$ and $\Gamma'$ be surfaces such that
$\gamma\circ\Gamma=\Gamma'$.
Suppose that there are two insertions
$\gamma_{_{K}}\circ\Gamma_J=\Gamma'_I$ and
$\gamma_{_K}\tilde{\circ}\Gamma_{\sigma(J)}=\Gamma'_{\sigma'(I)}$,
where
$I=\{1,\:\ldots,\:P_{\Gamma'}\}$,
$J=\{1,\:\ldots,\:P_{\Gamma}\}$ and
$K=\{P_{\Gamma}+1,\:\ldots,\:P_{\Gamma'}\}$.
Assume that there exists
$j:\;\Gamma_J\rightarrow\Gamma_{\sigma(J)}$
and there is no
$j':\;\Gamma'_I\rightarrow\Gamma'_{\sigma'(I)}$.
Thus we have a new symmetry in $\Gamma$ after the extraction of $\gamma$ from $\Gamma'$.
We denote by $N_C(\gamma,\Gamma;\Gamma')$ the number of such symmetries.
This number is equal to the number of all inequivalent insertions $\gamma\circ\Gamma$.
If $\Gamma'=\gamma\cdot\Gamma$, then we define $N_C(\gamma,\Gamma;\Gamma')=1$.

Consequently the number of symmetries
$$
Z(\Gamma)
=Z(\bar{\gamma}_1)\;\frac{Z(\Gamma/\gamma_1)}{N_C(\bar{\gamma}_1,\Gamma/\gamma_1;\Gamma)}
N(\bar{\gamma}_1,\:\Gamma/\gamma_1;\:\Gamma)
$$
or
$$
Z(\Gamma)N_C(\Gamma_1,\Gamma_2;\Gamma)
=Z(\Gamma_1)Z(\Gamma_2)\;N(\Gamma_1,\Gamma_2;\Gamma).
$$

The new basis is $\widetilde{S^i}=Z(S_i)S^i$.
%For brevity, we define the product of 1PI surfaces $\widetilde{S^i}$ and $\widetilde{S^j}$.
Let $S^i$ and $S^j$ be 1PI surfaces, then the product
$$
\begin{array}{lll}
Z(S_i)\:S^i* Z(S_j)\:S^j
&=&\sum_k Z(S_i)Z(S_j)\;N(S_i,\:S_j;\:S_k)S^k
\medskip\\
&=&\sum_k Z(S_k)\;N_C(S_i,S_j;S_k)S^k
\medskip\\
&=&Z(S_i)Z(S_j)\;S^i\cdot S^j+
\sum^{\not\sim}_{s^i\circ s^j}Z(S_i\circ S_j)\;S^i\circ S^j,
\end{array}
$$
the last sum is over all
inequivalent insertions $S^i\circ S^j$,
the surface $S^i\cdot S^j$ is a disjoint union of $S^i$ and $S^j$.
We have
$$
\widetilde{S^i}*\widetilde{S^j}
=\sum^{\not\sim}_{s^i\circ s^j}\widetilde{(S^i\circ S^j\:)}+\widetilde{S^i}\cdot\widetilde{S^j},
$$
where the notation $\widetilde{(S^i\circ S^j\:)}$ means
$Z(S_i\circ S_j)\;S^i\circ S^j$.
\medskip

{\Ap \label{Nc} Negative character.}

We have to prove that the function $C$ defined by (\ref{BR})
is a character,
i.e. $C(\Gamma_1\cdot\Gamma_2)=C(\Gamma_1)C(\Gamma_2),\;\forall\Gamma_1,\:\Gamma_2$.}

As in Appendix 2, we prove the proposition by
induction on the number of internal lines.
If $P_{\Gamma_1\cdot\Gamma_2}=0$, then $\Gamma_1=\Gamma_2=\emptyset$
and $C(\Gamma_1\cdot\Gamma_2)=C(\Gamma_1)C(\Gamma_2)=1$.

Assume that we have proved the assertion for all
$\gamma_1$ and $\gamma_2$ such that $P_{\gamma_1\cdot\gamma_2}\leq N$.
We have to prove it for $\Gamma_1$ and $\Gamma_2$ such that
$P_{\Gamma_1\cdot\Gamma_2}=N+1$.
If $\Gamma_1=\emptyset$ or $\Gamma_2=\emptyset$
then the proof is trivial.
Thus we assume that $\Gamma_1\neq\emptyset$ and
$\Gamma_2\neq\emptyset$.

By definition (\ref{BR})
$$
C(\Gamma)
=-P_{-}\left(F(\Gamma)
+{\sum_{\gamma\subset\Gamma}}'\;C(\bar{\gamma})\cdot F(\Gamma/\gamma)
\right)
=-P_{-}\left({\sum^{\neq}_{\gamma\subset\Gamma}}\;C(\bar{\gamma})\cdot F(\Gamma/\gamma)
\right),
$$
the last sum is over $\gamma\subset\Gamma,\;\bar{\gamma}\neq\Gamma$.
Consequently
$$
{\sum^{\neq}_{\gamma\subset\Gamma}}\;C(\bar{\gamma})\cdot F(\Gamma/\gamma)
=-C(\Gamma)+C_+(\Gamma),
$$
where $C_+(\Gamma)$ is a positive function.

Now
$$
\begin{array}{lll}
0&=&P_-(C* F)(\Gamma_1\cdot\Gamma_2)
\medskip\\
&=&P_-\left\{(C\otimes F)(\Delta\Gamma_1\cdot\Delta\Gamma_2)
\right\}
\medskip\\
&=&P_-\left\{(C\otimes F)
\left(\Gamma_1\otimes\emptyset+
{\sum^{\neq}_{\gamma_1\subset\Gamma_1}}\bar{\gamma}_1\otimes\Gamma_1/\gamma_1\right)\right.\\
&&\left.\left(\Gamma_2\otimes\emptyset+
{\sum^{\neq}_{\gamma_2\subset\Gamma_2}}\bar{\gamma}_2\otimes\Gamma_2/\gamma_2\right)
\right\}
\medskip\\
&=&P_-\left\{C(\Gamma_1\cdot\Gamma_2)\right.\\
&&\left.
+{\sum^{\neq}_{\gamma_2\subset\Gamma_2}}C(\Gamma_1\cdot\bar{\gamma}_2)F(\Gamma_2/\gamma_2)\right.\\
&&\left.
+{\sum^{\neq}_{\gamma_1\subset\Gamma_1}}C(\Gamma_2\cdot\bar{\gamma}_1)F(\Gamma_1/\gamma_1)\right.\\
&&\left.+{\sum^{\neq}_{\gamma_1\subset\Gamma_1}}{\sum^{\neq}_{\gamma_2\subset\Gamma_2}}
C(\bar{\gamma_1}\cdot\bar{\gamma_2})F(\Gamma_1/\gamma_1\cdot\Gamma_2/\gamma_2)
\right\}
\medskip\\
&=&P_-\left\{C(\Gamma_1\cdot\Gamma_2)\right.\\
&&\left.+C(\Gamma_1)(-C(\Gamma_2)+C_+(\Gamma_2))\right.\\
&&\left.+C(\Gamma_2)(-C(\Gamma_1)+C_+(\Gamma_1))\right.\\
&&\left.+(-C(\Gamma_1)+C_+(\Gamma_1))(-C(\Gamma_2)+C_+(\Gamma_2))\right\}
\medskip\\
&=&P_-\left\{C(\Gamma_1\cdot\Gamma_2)-C(\Gamma_1)C(\Gamma_2)
+C_+(\Gamma_1)C_+(\Gamma_2)
\right\}
\medskip\\
&=&C(\Gamma_1\cdot\Gamma_2)-C(\Gamma_1)C(\Gamma_2).
\end{array}
$$
Consequently
$$
C(\Gamma_1\cdot\Gamma_2)=C(\Gamma_1)C(\Gamma_2),
$$
and the assertion is proved.

\bigskip
\bigskip
{\large\bf Acknowledgements.}

\bigskip
The author is grateful to Alexey Morozov and Konstantin Selivanov
for enormous help on all stages of this project and
to S.~Klevtsov, T.~Dymarsky, K.~Trushkin and R.~Anno for useful discussions.
%Special thanks to my father for advises and attention to the work.
Special thanks to A.~Mironov and S.~Rychkov for careful reading of the
manuscript.
The work is partially supported by RFBR grant 00-02-16477
and Russian president's grant 00-15-99296.

\end{document}